\documentclass[lettersize,journal,twoside]{IEEEtran}
\usepackage{amsmath,amssymb,amsfonts}
\usepackage{algorithmic}
\usepackage[ruled]{algorithm2e}
\usepackage{array}
\usepackage{textcomp}
\usepackage{stfloats}
\usepackage{url}
\usepackage{color}
\usepackage{verbatim}
\usepackage{graphicx}
\usepackage{tabularx}
\usepackage{multirow}
\usepackage{array}
\usepackage{booktabs}
\usepackage{graphicx}
\usepackage{parskip}
\newtheorem{theorem}{Theorem}
\newtheorem{definition}{Definition}
\usepackage[numbers]{natbib}
\usepackage{bm}
\usepackage{mathtools}
\DeclareMathOperator*{\argmax}{arg\,max}
\usepackage{subfigure}
\usepackage{bibspacing}
\graphicspath{ {./fig/} }

\def\BibTeX{{\rm B\kern-.05em{\sc i\kern-.025em b}\kern-.08em
    T\kern-.1667em\lower.7ex\hbox{E}\kern-.125emX}}
\usepackage{balance}

\begin{document}
	
\title{Hypergraph-Aided Task-Resource Matching for Maximizing Value of Task Completion in Collaborative IoT Systems}
\author{Botao~Zhu,~\IEEEmembership{Member,~IEEE} and Xianbin~Wang,~\IEEEmembership{Fellow,~IEEE}
\thanks{
This work was supported in part by the Discovery Program of Natural Sciences and Engineering Research Council of Canada (NSERC) under Grant
RGPIN-2024-05720 and in part by the Canada Research Chair Program. (\textit{Corresponding author: Xianbin Wang.})

B. Zhu and X. Wang are with the Department of Electrical and Computer Engineering, Western University, London, N6A 5B9, Canada (e-mail: bzhu88@uwo.ca; xianbin.wang@uwo.ca).
}}


\maketitle

\begin{abstract}
With the growing scale and intrinsic heterogeneity of Internet of Things (IoT) systems, distributed device collaboration becomes essential for effective task completion by dynamically utilizing limited communication and computing resources. However, the separated design and situation-agnostic operation of computing, communication and application layers create a fundamental challenge for rapid task-resource matching, which further deteriorate the overall task completion effectiveness. To overcome this challenge, we utilize hypergraph as a new tool to vertically unify computing, communication, and task aspects of IoT systems for an effective matching by accurately capturing the relationships between tasks and communication and computing resources. Specifically, a state-of-the-art task-resource matching hypergraph (TRM-hypergraph) model is proposed in this paper, which is used to effectively transform the process of allocating complex heterogeneous resources to convoluted tasks into a hypergraph matching problem. Taking into account computational complexity and storage, a game-theoretic hypergraph matching algorithm is proposed via considering the hypergraph matching problem as a non-cooperative multi-player clustering game. Numerical results demonstrate that the proposed TRM-hypergraph model achieves superior performance in matching of tasks and resources compared with comparison algorithms.  
\end{abstract}


\section{Introduction}
	
\IEEEPARstart{T}HE proliferation of diverse Internet of Things (IoT) devices, coupled with variations in their operating environments, has resulted in ubiquitous heterogeneity in the computing and communication resources available across IoT systems~\cite{9099866}. Furthermore, the stringent nature of tasks generated by various IoT platforms gives rise to dynamic and demanding requirements for computing and communication resources~\cite{9542866}. The increasing complexity of tasks and resources presents a fundamental challenge in allocating the appropriate resources to tasks, which further result in deteriorated resource utilization. Therefore, a paradigm shift is currently unfolding within the IoT system operation, marked by a distinct emphasis on a task-centric approach. This approach entails the precise and rapid allocation of communication and computing resources within the IoT system, ensuring their alignment with the distinctive task-centric requirements.

Most IoT devices have very limited computing and communication resources, posing a significant challenge for individual devices to cope with the ever-growing complexity of tasks~\cite{9869705}. Due to their diverse capabilities, collaboration among IoT devices plays a vital role in enabling the IoT system for accomplishing complex goals and tasks~\cite{8960404}. Through seamless collaboration, an IoT system is able to pool the distributed resources from IoT devices, exchange information, and coordinate their actions, ultimately achieving the system’s goal with enhanced task completion efficiency, accuracy, and outcomes~\cite{9091902}.

Collaborative computing in IoT, as an emerging technology, has attracted significant research interest for its ability to leverage the scattered computing and storage resources of IoT devices to provide high-bandwidth, low-latency services for resource demanding tasks~\cite{9999479}. In the existing studies, most researchers focus on resource allocation, task partitioning, and task scheduling~\cite{9674824} in the collaborative computing IoT system to achieve the ultimate goals, such as minimum energy consumption~\cite{7161307}, minimum delay~\cite{9477600}, and maximum quality of service (QoS)~\cite{9566207}. However, the successful completion of a collaborative task depends on many factors holistically, such as device, computation, communication, and task allocation. Therefore, the design of a collaborative system should be a comprehensive consideration of all collaboration-related factors. So far, crucial aspects for orchestrated operation have yet to be thoroughly addressed in the implementation of the collaborative computing IoT system.

One fundamental aspect needing more attention is the discovery of resources within the deployed IoT system. Resource discovery encompasses the procedures of identifying and locating IoT devices, ascertaining their computing capabilities, and assessing the quality of communication links between these devices~\cite{9225133}. In a collaborative computing system, if the discovery of collaborative resources for task execution happens after the task has been generated, it may lead to significant delay for task completion due to the significant amount of time on resource discovery. This delay makes attaining the low-latency requirement of the collaborative computing IoT system extremely challenging~\cite{9583910}.

Matching tasks with appropriate resources is another pivotal aspect in the collaborative computing IoT system due to the inherent heterogeneity of both elements. IoT devices vary in form, each possessing unique characteristics, functionalities, and specifications~\cite{9711564}. Similarly, tasks generated across various IoT platforms are diverse, varying in size and requiring distinct processing capabilities~\cite{9858872}. If the collaborative resource allocated by the IoT system fails to handle the task adequately or if the task's completion time surpasses the set deadline, it directly reduces utilization efficiency of allocated resource. Consequently, the resource initially assigned becomes ineffective and is unable to achieve its designated function. Hence, achieving the efficient matching between resources and tasks is essential for the optimal execution of tasks and the maximization of resource utilization.

In addition to the discovery of collaborative resources and the matching of resource-task pairs, coordinating the computing, communication, and application layer systems to achieve collaborative tasks poses another significant challenge due to the independent operation of these three systems within the IoT system. At the computing layer, IoT devices are equipped with powerful processors and embedded software, enabling them to perform complex data analysis, execute algorithms, and make intelligent decisions locally \cite{9951051}. The communication layer of the IoT system consists of a diverse array of communication protocols that allow devices to establish connections and exchange data. At the application layer, IoT devices run specialized software that enables them to perform specific tasks and provide valuable services to users \cite{9933027}. Even though the independence of these layers brings advantages such as improved efficiency, the challenge arises when these layers need to collaborate efficiently to achieve common objectives. Coordinating data sharing, resource allocation, and task execution across multiple layers becomes a complex endeavor, especially when devices operate in diverse and dynamic environments. In fact, it is a daunting challenge to represent the complex relationships among computing, communication, and task layers in IoT using traditional graphs. The complexity and multidimensionality of the interactions within these layers requires more sophisticated approaches to accurately capture their interdependencies.

Recent studies have shown that hypergraphs are the powerful tool for effectively representing multiple relationships of complext networks \cite{9264674}. Hypergraphs provide a more expressive and flexible representation, which allow multiple nodes to be connected to a single hyperedge. This enables a richer depiction of the complex relationships between different elements in the IoT system. By using hypergraphs, the interplay between computing resources, communication links, and task allocations in IoT can be effectively modeled. Hyperedges in hypergraphs can represent the combinations of computing, communication, and task, and provide a comprehensive view of how these layers interact and collaborate to achieve specific goals. In addition, hypergraphs facilitate the representation of higher-order relationships that are important for collaborative decision-making to optimize resource allocation and achieve efficient outcomes. Due to the growing complexity of the IoT system, utilizing hypergraphs to describe the interrelationships among the computation, communication, and task layers becomes crucial for gaining deeper insights into the system's behavior and improving its overall performance.

Taking into account the aforementioned challenges and the advantages of hypergraphs, in this research we propose a state-of-the-art task-resource matching hypergraph (TRM-hypergraph) model to implement the precise matching between tasks and resources in the collaborative computing IoT system, aiming at maximizing the value of task completion. The proposed TRM-hypergraph model
is the first time to leverage the advantages of hypergraphs for resource allocation in IoT systems,
providing a novel perspective and solution for this research field. The main contributions are summarized as follows:

\begin{itemize}
  \item
  Computing layer, communication layer, and task layer are considered holistically to coordinate resource allocation. By introducing the concepts of resource hypergraph and task hypergraph,
  hypergraphs are utilized to integrate  computing, communication, and task layers by accurately capturing the cooperative relationships among intricate tasks, computing resources, and communication resources within the collaborative computing IoT system.

  \item The TRM-hypergraph model is proposed to achieve the precise resource allocation mechanism tailored to different task requirements. Within this model, the challenge of resource discovery and allocation is transformed into the hypergraph matching problem between the task hypergraph and the resource hypergraph.

  \item The converted hypergraph matching problem is solved through a game theory perspective, introducing a game-theoretic approach that treats the problem as a non-cooperative multiplayer clustering game. Simulation results are supplemented to illustrate that the optimal value of task completion is obtained by our proposed TRM-hypergraph model.
\end{itemize}

The rest of the paper is organized as follows. The literature review of related work is discussed in Section \ref{Related Work}. The system model and problem formulation are introduced in Section \ref{system model} and Section \ref{problem formulation}, respectively. The TRM-hypergraph model is presented in Section \ref{matching model}, and the game-theoretic hypergraph matching algorithm is proposed in Section \ref{game-theoretic matching}. Simulation results are provided in Section \ref{section simulation}. Finally, Section \ref{conclusion} concludes this paper. 

\section{Related Work}
\label{Related Work}
In this section, a comprehensive review of existing research in the field of collaborative computing is conducted, and the significance of our work in enhancing the collaborative computing system is elucidated. Then, the primary application domains of hypergraphs are explored.

\subsection{Application of Collaborative Computing in Wireless Networks}
 Collaborative computing system has attracted increasing attention from industry and academia in recent years because each device in the system can contribute its processing power, memory, and storage to work collectively on a task. In \cite{7161307}, the authors proposed a cooperation node selection strategy to achieve a trade-off between fairness and energy consumption at each node. The authors in \cite{8674597} presented a reliability-oriented cooperative computation framework in the cooperative vehicle-infrastructure system to  maximize the coupled reliability of communication and computation. In \cite{9764636}, the authors minimized the weighted sum of latency and energy consumption by jointly optimizing offloading selection, computing frequency control, and transmission power control in the device-to-device-enabled multi-helper terminal cooperative computing network. In \cite{8798727}, the authors designed an online learning-aided cooperative offloading mechanism by considering computation, transmission, and trust services to minimize the loss. By taking into account computation offloading selection, clock frequency control, and transmission power allocation, an energy-efficient dynamic offloading and resource scheduling policy is developed to reduce energy consumption in the mobile cloud computing system in \cite{8352664}. In \cite{9447211}, the authors proposed a situation-aware offloading scheme in the collaborative computing system to maximize value of service. In \cite{9290426}, the authors proposed a deep reinforcement learning-based algorithm for joint server selection, cooperative offloading, and handover in a multi-user edge wireless network to minimize computation cost in terms of total delay. The aforementioned works implemented cooperative computing in various scenarios. However, they have not considered the diversity of computing resources and tasks and the matching of tasks and resources. Hence, this research will focus on achieving efficient matching between complex tasks and intricate resources via building the relationships between them.

\subsection{Utilizing Hypergraphs to Depict Higher-order Relationships}

Due to the advantages of hypergraphs in representing higher-order relationships, hypergraphs gain significant research attention in the fields of image processing and computer vision. In \cite{9292465}, the authors proposed the spectrum-based hypergraph construction methods for both clean and noisy point clouds, which improve the performance in sampling and denoising applications. To solve the problem of hyperspectral image classification, the authors in \cite{10196337} developed the dynamic hypergraph convolution and recursive gated convolution fusion network, which can dynamically update the hypergraph model and capture the global spatial information of the hyperspectral image. The authors in \cite{10049798} proposed a hypergraph parser to simulate guiding perception to study intra-modal object-wise relationships for solving the problem of multimodal remote sensing image segmentation. In \cite{9788570}, the authors presented the hypergraph construction-compression-conversion to detect 3D objects from noise point clouds, which can capture the variances from multiple features and improve the graph representations in point clouds. In \cite{8478794}, the authors used the hypergraph-induced convolutional network to understand the high-order correlation in visual data in which the high-order correlation is optimized in a learning process. {Compared with the commonly used graph-based methods in the fields of image and computer vision, the hypergraph-based methods mentioned above show superior performance in exploring high-order information and mining latent data relationships.}

\begin{table}[!t]
	\centering
        \renewcommand{\arraystretch}{1.2}
	\caption{Summary of Key Notations}
	\label{notation}
	  \begin{tabular}{p{1cm}<{}p{7cm}<{}}
		\hline \textbf{Notation} & \textbf{Description} \\
		\hline
		  {$a_j$} & {A device, $a_j \in \bm{A}$.} \\
             {$\bm{A}$} & {The set of all devices.} \\
		   {$\bm{B}$} & {Task.} \\
              {$b_m$} & {A subtask, $b_m \in \bm{B}$.}\\
              {$o_s$} & {A task type.} \\
              {$\bm{O}_{a_j}$} &  {The set of task types supported by $a_j$}\\
              {$r_{a_j,a_i}$} & {The average transmission rate from $a_j$ to $a_i$.} \\
              {$t_{b_m}$} & {The completion time of subtask $b_m$.} \\
              {$E_{b_m}$} & {The total energy consumption for completing $b_m$.}\\
              {$V_{b_m}$} & {Value of task completion.}\\
              {$\mathcal{H}^{\text{re}}$} & {The resource hypergraph.}\\
              {$\mathcal{V}^{\text{re}}$} & {The set of all involved objects in $\mathcal{H}^{\text{re}}$.} \\
              {$|\mathcal{V}^{\text{re}}|$} & {The number of objects in $\mathcal{H}^{\text{re}}$.} \\
              {$\mathcal{E}^{\text{re}}$} & {The set of hyperedges in $\mathcal{H}^{\text{re}}.$} \\
              {$e^{\text{re}}_{a_j,o_s,a_i}$} & {A resource hyperedge in $\mathcal{E}^{\text{re}}$.} \\
              {$\mathcal{H}^{\text{ta}}$} & {The task hyperegraph.}\\
              {$\mathcal{V}^{\text{ta}}$} & {The set of all involved objects in $\mathcal{H}^{\text{ta}}$.}\\
              {$|\mathcal{V}^{\text{ta}}|$} & {The number of objects in $\mathcal{H}^{\text{ta}}$.} \\
              {$\mathcal{E}^{\text{ta}}$} & {The set of all hyperedges in $\mathcal{H}^{\text{ta}}$. }\\
              {$e^{\text{ta}}_{o_s}$} & {The task hyperedge formed by a subtask with type $o_s$ and the task initiator $a_j$, $e_{o_s} \in \mathcal{E}^{\text{ta}}$.}\\
              {$\bm{\mathcal{F}}$} & {The set of all potential assignment matrices that map each vertex of $\mathcal{V}^{\text{ta}}$ to each vertex of $\mathcal{V}^{\text{re}}$.} \\ 
              {$\bm{f}^*$} & {The assignment matrix that can maximize the overall matching score, $\bm{f}^* \in \bm{\mathcal{F}}$.}\\
              {$\bm{P}$} & {The set of players. }\\
              {$\bm{N}$} & {The set of strategies, $\bm{N}=\{1,\dots,N \}$. }\\
              {$\bm{q}$} & {The state of population.} \\
              {$q_n$} & {The fraction of players selecting the $n$-th strategy.}\\
            \hline
   \end{tabular}
\end{table}

{In addition to image processing and computer vision, there have been several studies in recent years that utilize hypergraphs to optimize resource allocation in wireless networks, particularly in IoT. In \cite{10268966}, the authors proposed the hypergraph-based resource-efficient collaborative reinforcement learning for beyond 5G massive IoT to improve resource utilization. In \cite{9813716}, the authors converted the channel resource allocation issue in industrial IoT to the maximum weighted clique set problem in hypergraph theory, and solved this problem by proposing the hypergraph spectral clustering method. The authors in \cite{8647400} presented a hypergraph-based 3D matching method to jointly optimize relay selection, channel allocation, and power control to maximize energy efficiency in IoT. The authors in  \cite{10458122} developed a hypergraph interference model and two reinforcement learning-based resource management algorithms for the 6G-enabled massive IoT to enhance the network throughput.} {In \cite{8469018}, the authors propose a hypergraph spectral clustering-based algorithm to deal with the strong interference and severe cumulative interference in dense NOMA-HetNet.} In \cite{7583644}, the authors proposed a hypergraph-based learning solution to solve the distributed channel access problem in device-to-device communication. To evaluate the clustered cooperative beamforming in cellular networks, the authors in \cite{7329952} presented a framework based on a hypergraph formalism to achieve the optimal use of cache and backhaul resources in cellular networks. {The aforementioned studies utilize hypergraphs to describe the relationships among complex resources, thereby achieving improved system performance compared to the contrast algorithms employed in their research.}

The above analysis highlights the significant advantages of hypergraphs in various domains. This research delves into the incorporation of hypergraphs into the collaborative computing IoT system to enhance overall system performance.

\section{Modeling of Collaborative Computing IoT System and Evaluation Criteria for Task Completion}
\label{system model}

\begin{figure}[!]
 \centering
 \includegraphics[scale=0.58]{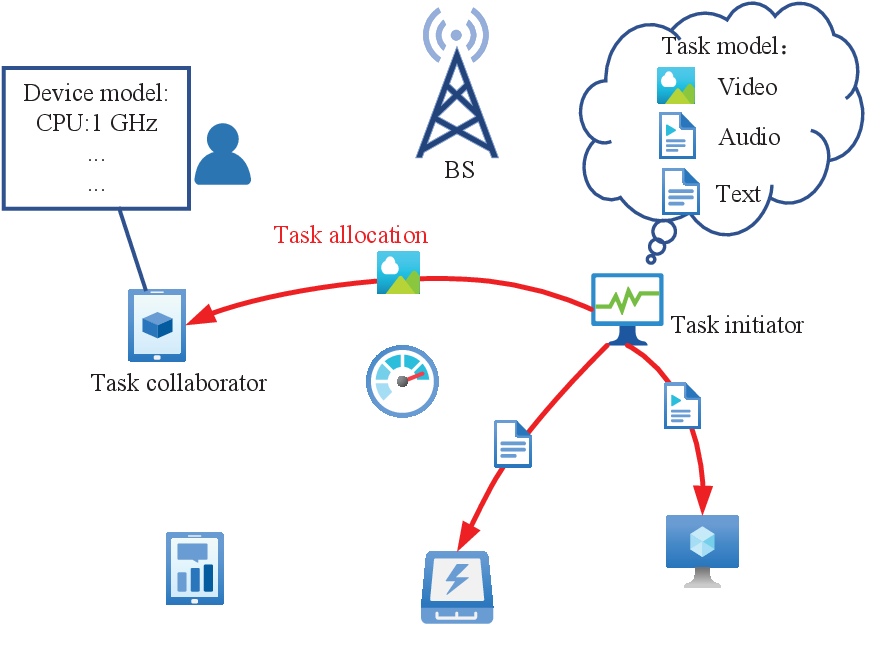}
 \caption{A collaborative computing IoT system.}
 \label{collaborative computing}
 \end{figure}

The architecture of a collaborative computing IoT system is illustrated in Fig.~\ref{collaborative computing}. The base station (BS) is responsible for collecting data, such as location information, energy, etc., from IoT sensors or terminal devices connected with it. Each device can be both a task initiator and a task collaborator\footnote{In this paper, the collaborator and collaborative computing device are used interchangeably.}, which is determined by its state. We assume the set of all devices in the considered system is $\bm{A} =\{a_1,a_2, \dots, a_J\}$. {The purpose of collaborative computing is to accomplish tasks more efficiently. The successful matching of collaborators and tasks primarily depends on the characteristics of the available physical resources and tasks. Therefore, to better elucidate the proposed TRM-hypergraph, the task model, the physical resource model, and the task execution model are first accurately defined and described in this section. Additionally, to assess the effectiveness of resource and task matching, the value of task completion is also elaborately introduced. The main notations used in this research are summarized in Table~\ref{notation}.}

\subsection{Task Model}
Any device can generate a task $\bm{B}$, which is composed of a set of subtasks $\{b_1, b_2,\dots, b_M \}$. These subtasks vary in terms of size, processing density, and task type. For example, a streaming media task includes video processing, voice processing, and subtitle processing subtasks. Consequently, distinct computing resources are essential for separately handling these subtasks. A model that fully encompasses all aspects of a task is inherently complex. It requires careful consideration of numerous details. However, incorporating an excessive level of detail can make the problem mathematically intractable, without providing practical insights for engineering purposes. In this research, 
the subtask $b_m, m=1,\dots, M$, is characterized as $I_{b_m}= (d_{b_m}, \rho_{b_{m}}, o_s, t^{\text{max}}_{b_m})$: 
\begin{itemize}
  \item \textit{Data size} $d_{b_m}$: The number of data bits of $b_m$.
  \item \textit{Processing density} $\rho_{b_m}$: Each computing task has a processing density, depending on the type of task, which is defined as required CPU cycles to process a unit bit (cycles/bit) \cite{7264984}.   
  \item \textit{Task type} $o_s$: It is the task type of $b_m$. In $\bm{B}$, each subtask corresponds to a different task type. All task types in the considered IoT system are defined as $\bm{O} = \{o_s\}_{s=1}^S$. 
  \item \textit{Task completion time} $t_{b_m}^{\text{max}}$: It is the maximum task completion tolerance time, determined by the task initiator at the initiation of the task.
\end{itemize}

It is worth pointing out that all subtasks are executed independently, and there is no mutual dependency between subtasks. Finally, the parameterized task information $\bm{I} = \{I_{b_1}, \dots, I_{b_M}\} $ is obtained.

\subsection{Physical Resource Model}

The physical resource model is the most crucial component of the system model because it determines what tasks each device can perform and how well it supports tasks. The physical resource model considered in this research includes the device resource model and the communication link model, which are described in detail as follows.

\subsubsection{Device Resource Model}
Due to the significant diversities of IoT devices, such as CPU, buffer, storage, power source, cost, etc., it is challenging to find a model that accurately represents the characteristics of IoT devices. Without loss of generality, all devices are equipped with a single-core CPU and a single antenna, and they have sufficient energy for transmitting, receiving, and computing the data. By capturing the main features of each device, we parameterize device $a_j, j=1,\dots, J$, to be $a_j = (id_{a_j}, f_{a_j}, g_{a_j}, \bm{O}_{a_j})$:
\begin{itemize}
    \item Device ID $id_{a_j}$: It is the unique identifier for each device. 

    \item CPU clock frequency $f_{a_j}$: CPU clock frequency is the number of cycles a CPU performs per second. CPUs have a base clock frequency and a boost clock frequency. The base clock frequency represents the default operating frequency of the CPU, while the boost clock frequency indicates the CPU's ability to dynamically increase its speed on a single core when encountering demanding workloads. However, consistently running the CPU at boost clock frequency poses various challenges, including excessive heat generation, increased power consumption, and potential risks of CPU failure. To simplify the problem, the CPU of each device is assumed to be operated at a constant base clock frequency. 



    \item Coordinate $g_{a_j}$: It is the coordinate of device $a_j$ in 3D space.


    \item Set of task types $\bm{O}_{a_j}$: A set of task types supported by device $a_j$, which is the subset of $\bm{O}$. 
   
\end{itemize}

\subsubsection{Communication Link Model} Each device can establish a link with another device with the assistance of the BS. If device $a_j$ can establish the communication link with device $a_i$, then the achievable average transmission rate $r_{a_j, a_i}$ is given by
\begin{align}
    r_{a_j, a_i} = W\log_2\left(1 +  \gamma_{a_j, a_i} \right), \gamma_{a_j, a_i} \geq \gamma_0,
\end{align}
where $W$ and $\gamma_{a_j, a_i}$ are the bandwidth and the signal-to-noise ratio (SNR) of the transmission link, respectively. $\gamma_{a_j, a_i} = \frac{p_{a_j}h_{a_j,a_i}}{N_o}$, where $h_{a_j,a_i}$ is the channel gain, $N_0$ is the noise power, and $p_{a_j}$ is the transmission power which is determined by the BS according to power control algorithms \cite{7307234}. $\gamma_0$ is the SNR threshold and the transmission is considered successful if the SNR is greater than the threshold, which is used to meet the minimal transmission requirement. A simple channel model $h_{a_j,a_i} = |g_{a_j}-g_{a_i}|^{-\alpha_0}$ is considered in this research where $|g_{a_j}-g_{a_i}|$ is the distance between $a_j$ and $a_i$, and $\alpha_0=4$ is the path loss factor \cite{9756368}. If device $a_i$ can transmit data to device $a_j$, the achievable average transmission rate is $r_{a_i, a_j}$, which may not be equal to $r_{a_j, a_i}$.

\subsection{Task Execution Model}
If device $a_j$, as a task initiator, generates a task $\bm{B}$ that needs to be completed by other devices, it sends a resource request message with the task feature information $\bm{I}$ to the BS. According to the received request and available devices, the BS selects the most suitable devices as task collaborators and returns the results to $a_j$. After that,  $a_j$ establishes connections with the selected task collaborators and sequentially transmits subtasks to them. After each task collaborator executes the task, it will send the result to the task initiator. Since the energy consumption and time of communication between $a_j$ and the BS are usually very small, they are ignored in our considered system. In addition, the time and energy utilized in building links and transmitting task results between $a_j$ and task collaborators are also ignored. Therefore, the scenarios under consideration primarily involve the offloading of subtasks from $a_j$ to all task collaborators and the subsequent computation of these subtasks.

If the subtask $b_m$ with the task type $o_s$ is offloaded to $a_{i}$, then $a_i$ should fulfill the condition that $a_j$ can establish a valid link with it, and it supports $o_s$.
Hence, the transmission time of $b_m$ from device $a_j$ to $a_i$ is given by
\begin{align}
    t_{b_m}^{\text{tra}} =  \frac{d_{b_{m}}}{r_{a_j, a_i}}.
\end{align}
The energy consumed for transmitting $b_m$ is given by $E^{\text{tra}}_{b_m} = p_{a_j}t^{\text{tra}}_{b_m}$. After completing the transmission of $b_{m}$, device $a_j$ continues to offload the next subtask to another selected task collaborator, and device $a_i$ starts to execute the subtask $b_m$. The execution time of the subtask $b_m$ is
\begin{align}
   t_{b_m}^{\text{exe}} = \frac{d_{b_{m}}\rho_{b_{m}}}{f_{a_{i}}}. 
\end{align}
Hence, the completion time of subtask $b_m$ is $t_{b_m} = t_{b_m}^{\text{tra}} + t_{b_m}^{\text{exe}}$. The computational energy consumed by device $a_i$ for computing $b_m$ is \cite{6195685}
\begin{align}
\label{exe_energy}
    E^{\text{exe}}_{b_m} = \alpha_1 f^2_{a_i} d_{b_{m}}\rho_{b_{m}},
\end{align}
where $\alpha_1 f^2_{a_i}$ is the coefficient denoting the consumed energy per CPU cycle, and $\alpha_1$ is set to $10^{-11}$ according to the measurements in \cite{6195685}. Therefore, the total energy consumption for completing $b_m$ is given by $E_{b_m} = E^{\text{tra}}_{b_m} + E^{\text{exe}}_{b_m}$.

\subsection{Value of Task Completion as a Metric}
To evaluate the performance of task completion, an innovative metric, value of task completion, is proposed. Value has been extensively applied in the modeling of market activities, characterized as the user's holistic evaluation of the value derived from a product or service, taking into account user's perception. As users are typically unaware of the actual production or service costs, they will only make a purchase when their perceived value for the product surpasses its selling price. In our considered collaborative computing IoT network, what task initiator can directly perceive is the amount of energy and time spending on computing tasks. Therefore, the considered value of task completion comprises both the value of time and the value of energy.

\subsubsection{Value of time}
The value of time of completing the subtask $b_m$ is formulated as
\begin{align}
    \label{time value}
    V^{\text{time}}_{b_m} =
     e^{\frac{\left(t^{\text{max}}_{b_m} - t_{b_m}\right)}{t^{\text{max}}_{b_m}}},
\end{align}
where $t^{\text{max}}_{b_m}$ is the maximum task completion tolerance time, and $t_{b_{m}}$ is the actual completion time of $b_{m}$, which is determined by the selected task collaborator and communication link. If the actual completion time is equal to the maximal completion time, the value of time is set to 1. If $t_{b_{m}}$ is less than $t_{b_m}^{\text{exp}}$, $V^{\text{time}}_{b_m}$ is greater than 1 and increases as the actual completion time decreases.  

\subsubsection{Value of energy}
The value of energy of completing the subtask $b_{m}$ is defined as the comparison between the expected energy consumption and the actual energy consumption
\begin{align}
    \label{cost value}
    V^{\text{ener}}_{b_m} = 
      e^{\frac{(E^{\text{exp}}_{b_m} - E_{b_m})}{E_{b_m}^{\text{exp}}}},
\end{align}
where $E_{b_m}^{\text{exp}}$ and $E_{b_m}$ are the expected energy consumption and the actual energy consumption, respectively. Regarding the expected energy consumption $E_{b_m}^{\text{exp}}$, the subtask $b_{m}$ is assumed to be completed by the task initiator $a_j$ itself. The expected energy consumption is then calculated using the same method as described in equation~(\ref{exe_energy}).

\subsubsection{Value of task completion}

By considering values of time and energy, the value of task completion of the task initiator $a_j$ in completing the subtask $b_m$ is formulated as, 
\begin{align}
    \label{value}
     V_{b_m} = 
      \xi_1 V^{\text{time}}_{b_m} + \xi_2 V^{\text{ener}}_{b_m}, 
\end{align}
where $\xi_1$ and $\xi_2$ are the weight parameters of the value of time and the value of energy, respectively, $0 \leq \xi_1\leq 1$,  $0 \leq \xi_2\leq 1$. $\xi_1$ and $\xi_2$ provide rich model flexibility so that the system can flexibly consider the proportion of the value of time and the value of energy in making decisions to meet different needs.

\section{Problem Formulation}
\label{problem formulation}
According to (\ref{time value})--(\ref{value}), the value of task completion is determined by the selected task collaborator and communication link. {Therefore, to complete $\bm{B} = \{b_1, \dots, b_M \}$ generated by $a_j$, it is important to allocate the appropriate resources to all subtasks.} The allocation relationship between the subtask $b_m$ and the device $a_i$ is defined as $l_{b_m, a_i} = 1$ if $b_m$ is allocated to $a_i$, otherwise $l_{b_m, a_i} = 0$.

Our goal is to maximize the value of task completion in completing all subtasks of $\bm{B}$ through joint optimizing resource selection and allocation in the collaborative computing IoT system. The optimization problem is formulated as
\begin{alignat}{1}
    P1: &\max_{} \quad  \sum_{m=1}^{M}\sum_{i=1}^{J} l_{b_m, a_i} V_{b_m}   \\
    \mathrm{s.t.} \quad
    & \sum_{i=1}^{J}l_{b_m, a_i} = 1, m=\{1,\dots, M\}, \label{sub-1}\\
    & \sum_{m=1}^{M}l_{b_m,a_i} \leq 1, a_i \in \bm{A}, a_i\neq a_j, \label{sub-2}\\
    & o_s \in \bm{O}_{a_i}, \label{sub-3} \\ 
    & \gamma_{a_j, a_i} \geq \gamma_0. \label{sub-4}
\end{alignat}
Constraint (\ref{sub-1}) states that each subtask can be allocated to only one device, and constraint (\ref{sub-2}) declares that each device can be assigned at most one subtask. Constraint (\ref{sub-3}) specifies that the task type $o_s$ of subtask $b_m$ must be supported by $a_i$ if $b_m$ is assigned to $a_i$. Constraint (\ref{sub-4}) states that the SNR of the communication link between the task initiator $a_j$ and any collaborative device $a_i$ must exceed the threshold $\gamma_0$ to achieve successful task transmission. 

 \begin{theorem}
     \textit{Problem P1 is NP-hard.}
 \end{theorem}
\textit{Proof:} Assuming that the number of available devices equals the total number of subtasks and these devices support all task types in $\bm{B}$, 
the specific case of $P1$ involves the task initiator accessing each collaborative device, completing a subtask to obtain a value from that device, and repeating this process until all subtasks are completed. 
If each device is considered as a set, and the task types within each device are regarded as points in the set, then the collection of all sets of points in the system is represented as $\bm{\mathcal{O}} = \{\bm{O}_{a_1},\dots, \bm{O}_{a_M} \}$. $l_{k,c}=1$ represent the task initiator visits point $k \in \bm{\mathcal{O}}$ after visiting point $c \in \bm{\mathcal{O}}$, and $V_{k,c}$ is the obtained cost, which can be seen as the negative of the value of task completion. 
Hence, the special case of $P1$ is to minimize the total cost of visiting exactly one point from each set, which can be written as
\begin{align}
    &\min \quad \sum_{k=1}^{|\bm{\mathcal{O}}|} \sum_{c=1}^{\bm{|\mathcal{O}|}} l_{k,c} V_{k,c} \\
    \mathrm{s.t.} \quad
     & \sum_{k=1, k\neq c}^{|\bm{\mathcal{O}}|}l_{k,c} = 1, \forall c \in \bm{\mathcal{O}},\\
    & \sum_{c=1, c\neq k}^{|\bm{\mathcal{O}}|} l_{k,c} = 1, \forall k \in \bm{\mathcal{O}},\\
    &  l_{k,c} \in \{0, 1\}, \forall k,c \in \bm{\mathcal{O}},\\
    &  \sum_{\substack{k \in \bm{O}_{a_m}, \\ c \notin \bm{O}_{a_m}}  }l_{k,c} = 1, \forall c \in \bm{\mathcal{O}}, \forall \bm{O}_{a_m}  \in  \bm{\mathcal{O}}.
\end{align}
The special case is so-called the generalized traveling salesman problem (GTSP). Since  GTSP is NP-hard \cite{POP2024819}, $P1$ is hence also NP-hard.


\section{Task-Resource Matching Hypergraph Model}
\label{matching model}

To maximize the value of task completion in the collaborative computing system, it is crucial to consider not only the selection of available computing and communication resources but also the effective allocation between resources and subtasks. Since the computing layer, communication layer, and task layer are mutually independent, it is challenging to optimally select and allocate resources to accomplish tasks. Therefore, integrating the computing, communication, and task layers and capturing the relationships among these three layers to achieve precise tasks and resources matching is crucial in the collaborative system.

Obviously, the assignment relationships between tasks and resources cannot be accurately described using the traditional pairwise relationship representation. For example, if a communication link can be established between the task initiator $a_j$ and the task collaborator $a_i$, then a line can be used to connect them to represent their connection relationship. However, this connection line does not indicate whether $a_i$ is available or whether $a_j$ and $a_i$ support the same task types. To capture such complex relationships in the collaborative computing network, hypergraphs are employed in this research in lieu of conventional graphs. Hypergraphs allow the modeling of relations not only in pairs but also involving groups of vertices \cite{7299164}. Leveraging the benefits of hypergraphs, we introduce the cutting-edge TRM-hypergraph model for the collaborative computing system, aiming to facilitate resource-task matching. This comprehensive model consists of the collaboration-driven resource hypergraph, the task hypergraph, and the matching mechanism bridging the task hypergraph and the collaboration-driven resource hypergraph. The detailed description of this model is outlined below.

\begin{figure}[t!]
\centering
\includegraphics[scale=0.95]{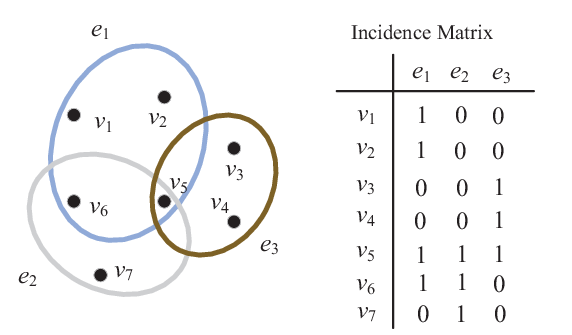}
\caption{A hypergraph with the incidence matrix representation.}
\label{hypergraph}
\end{figure}

\subsection{Preliminaries}

\subsubsection{Hypergraph} A hypergraph $\mathcal{H}$ is denoted by $\mathcal{H} = (\mathcal{V}, \mathcal{E})$, where $\mathcal{V}$ is the set of all vertices, $\mathcal{E}$ represents all the hyperedges. A heyperedge $e \in \mathcal{E}$ is a subset of $\mathcal{V}$. The incidence matrix is often utilized to represent hypergraphs, as shown in Fig.~\ref{hypergraph}. Each row of the incidence matrix corresponds to a vertex, and each column corresponds to a hyperedge. All vertices of the hypergraph $\mathcal{H}$ in Fig.~\ref{hypergraph} is the set $\{ v_1,v_2,v_3,v_4,v_5,v_6,v_7\}$. The hyperedges $e_1$, $e_2$, and $e_3$ are $\{v_1,v_2, v_5,v_6 \}$, $\{v_5, v_6, v_7 \}$, and $\{v_3, v_4, v_5 \}$, respectively.

\subsubsection{$K$-uniform hypergraph} A hypergraph in which each hyperedge is associated with exactly $K$ vertices.

\subsection{Collaboration-driven Resource Hypergraph}
\label{resource hypergraph}
Completing a collaborative task involves discovering and selecting computing and link resources, as well as allocating resources to subtasks. When a device joins the system, it transmits its detailed information, including location, computing resource, supported task types, and other relevant information, to the BS. Using the gathered information, hypergraphs are employed to represent the complex collaboration relationships among tasks, computing resources, and communication resources. Collaboration is typically executed with multiple objects and corresponding collaboration conditions, which are formalized as follows.

\begin{definition}[Collaboration] \textit{A collaboration in the collaborative computing IoT system is an action done by multiple objects under a certain condition.  Given the set of objects $\mathcal{V}$, object attributes $\mathcal{A}^{\mathcal{V}}$, and collaboration conditions $\mathcal{A}^{\mathcal{E}}$, a collaboration is defined as $e = (\mathcal{V}_e, \mathcal{A}^{\mathcal{V}}_e, \mathcal{A}^{\mathcal{E}}_e)$, including all involved objects $\mathcal{V}_e \subset \mathcal{V}$ with their attributes $\mathcal{A}^{\mathcal{V}}_e \subset \mathcal{A}^{\mathcal{V}}$, and collaboration conditions $\mathcal{A}^{\mathcal{E}}_e \subset \mathcal{A}^{\mathcal{E}}$.}
\end{definition}

\begin{figure}[!]
\centering
\includegraphics[scale=0.68]{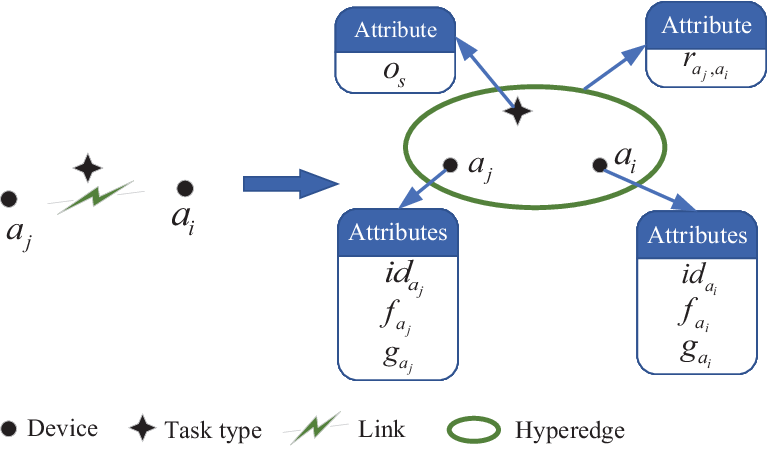}
\caption{An example of collaboration-driven hyperedge.}
\label{collaboration}
\end{figure}

 \begin{algorithm}[t!]
	\caption{{Generating collaboration-driven resource hypergraph}}
	\label{resource-hypergraph}
	\begin{algorithmic}[1]
		\renewcommand{\algorithmicrequire}{\textbf{Input:}}
		\REQUIRE $\bm{A}$
            \renewcommand{\algorithmicrequire}{\textbf{Output:}}
		\REQUIRE $\mathcal{H}^{\text{re}}$
             \STATE initialize an empty $\mathcal{H}^{\text{re}}$ and an empty graph $\mathcal{G}$\\
             \STATE randomly choose one device from $\bm{A}$ and add it to $\mathcal{G}$\\
             \WHILE{true}
                 \IF{all devices in $\bm{A}$ are visited}
                     \STATE \Return $\mathcal{H}^{\text{re}}$
                 \ENDIF
                 \STATE $\mathcal{G} \leftarrow$  randomly choose one new device $a_j$ from $\bm{A}$ \\ and connect with the device closest to it in physical distance in $\mathcal{G}$ \\
                 \STATE initialize an empty $Queue$ \\
                 \STATE label $a_j$ as explored \\
                 \STATE $Queue \leftarrow Queue.enqueue(a_j)$ \\
                 \WHILE{$Queue$ is not empty}
                     \STATE $a_i \leftarrow Queue.dequeue()$ \\
                      \FOR{$o_s$ in $\bm{O}_{a_j}$}
                          \FOR{$o_t$ in $\bm{O}_{a_i}$}
                              \IF{$o_s = o_t$}
                                  \IF{$a_j$ can establish link with $a_i$  $\&$ \\ $r_{a_j, a_i} \geq$ minimum transmission rate}
                     \STATE  create hyperedge $e^{\text{re}}_{a_j, o_s, a_i} = (a_j, o_s, a_i)$ \\ with the attributes of objects and the hyperedge weight $w^{\text{re}}_{a_j, o_s, a_i}$ \\
                     \STATE $\mathcal{H}^{\text{re}} \xleftarrow[]{add} {e^{\text{re}}_{a_j,o_s,a_i}}$
                   \ELSIF{$a_i$ can establish link with $a_j$  $\&$ \\ $r_{a_i, a_j} \geq$ minimum transmission rate}
                     \STATE  create hyperedge $e^{\text{re}}_{a_i,o_s,a_j} = (a_i, o_s, a_j)$ \\ with the attributes of objects and the hyperedge weight $w^{\text{re}}_{a_i,o_s,a_j}$\\
                     \STATE $\mathcal{H}^{\text{re}} \xleftarrow[]{add} {e^{\text{re}}_{a_i,o_s,a_j}}$
                         \ENDIF
                            \ENDIF
                          \ENDFOR
                      \ENDFOR

                     \FOR{all edges from $a_i$ to its adjacent devices}
                     \IF{adjacent device $a'$ is not labeled as explored}
                         \STATE label $a'$ as explored\\
                         \STATE $Queue \leftarrow Queue.enqueue(a')$ \\
                     \ENDIF
                   \ENDFOR
               \ENDWHILE 
             \ENDWHILE
	\end{algorithmic}
    \end{algorithm}
    
Take Fig. \ref{collaboration} as an example. If devices $a_j$ and $a_i$ support the task type $o_s$, $a_j$ can offload a subtask with type $o_s$ via a valid communication link to $a_i$ for collaborative computing. Hence, the collaboration involves three objects, including two devices (with attributes, such as $id_{a_j}$, $f_{a_j}$, and $g_{a_j}$) and a task (with attribute $o_s$), and a condition, i.e., link (with attribute $r_{a_j,a_i}$). To avoid the collaborative associations being split into several pairwise sub-relations as in traditional graphs, the attributed hypergraph is adopted to model the collaborations, which can preserve the collaborative associations among multiple objects. Specially, each collaboration is modeled as a hyperedge connecting multiple nodes that represent its involved objects. The attributes of objects are features of corresponding devices, and the weight of hyperedge is the link transmission rate. The collaboration-driven resource hypergraph in the collaborative computing system is formalized as follows.

\begin{definition}[Collaboration-driven Resource Hypergraph] \textit{ The collaboration-driven resource hypergraph is an attributed hypergraph $\mathcal{H}^{\text{re}} = \left(\mathcal{V}^{\text{re}}, \mathcal{E}^{\text{re}}, \mathcal{A}^{\text{re}}_{\mathcal{V}}, \mathcal{A}^{\text{re}}_{\mathcal{E}} \right)$, where $\mathcal{V}^{\text{re}}$ represents the set of all involved objects, including devices and task types, the hyperedge set $\mathcal{E}^{\text{re}}$ includes all the collaborations. $\mathcal{A}^{\text{re}}_{\mathcal{V}}$ and $\mathcal{A}^{\text{re}}_{\mathcal{E}}$ represent the feature matrices of objects and hyperedges, respectively. Since each hyperedge contains three objects, the constructed hypergraph is the 3-uniform hypergraph. }
\end{definition}

Based on the above modeling, the detailed process of generating the collaboration-driven hypergraph is presented in Algorithm \ref{resource-hypergraph}. The main idea of the algorithm is summarized as follows. When a new device joins the system, it traverses all existing devices in the system. If the new device can establish a valid communication link with one of the existing devices and they support the same task type, then a collaboration based on this task type is established between the two devices. Subsequently, a hyperedge containing this collaboration is generated and added to the hypergraph. For the consistency of notation, all objects involved in $\mathcal{V}^{\text{re}}$ are reformulated as $\mathcal{V}^{\text{re}} = \{v_1^{\text{re}}, \dots, v^{\text{re}}_{|\mathcal{V}^{\text{re}}|} \}$, where $|\mathcal{V}^{\text{re}}|$ is the number of objects. The set of hyperedges is reformulated as $\mathcal{E}^{\text{re}} = \{e_1^{\text{re}}, \dots, e^{\text{re}}_{|\mathcal{E}^{\text{re}}|} \}$, where $|\mathcal{E}^{\text{re}}|$ is the number of hyperedges.

\begin{algorithm}[h]
	\caption{{Generating task hypergraph}}
	\label{task-hypergraph}
	\begin{algorithmic}[1]
		\renewcommand{\algorithmicrequire}{\textbf{Input:}}
		\REQUIRE $\bm{B} = \{b_1, \dots, b_M \}$, $a_j$
		\renewcommand{\algorithmicrequire}{\textbf{Output:}}
		\REQUIRE Hypergraph $\mathcal{H}^{\text{ta}}$
		\STATE initialize an empty $\mathcal{H}^{\text{ta}}$\\  
		\FOR{$b_m$ in $\bm{B}$}
		\STATE $o_s \leftarrow b_m$ \\
		\STATE create hyperedge $e^{\text{ta}}_{o_s} = (a_j, o_s, \varnothing)$ with the object attributes $d_{b_m}, \rho_{b_m}, t_{b_m}^{\text{max}}$ and the hyperedge weight $w^{\text{ta}}_{o_s}$ \\
		\STATE $\mathcal{H}^{\text{ta}} \xleftarrow[]{add} {e^{\text{ta}}_{o_s}}$
		\ENDFOR
		\RETURN $\mathcal{H}^{\text{ta}}$
	\end{algorithmic}
\end{algorithm}

\subsection{Task Hypergraph}
\label{task hypergraph}
In the considered system, each task is associated with a task initiator, indicating that no task can exist independently of a device. Since any task initiator can generate a task $\bm{B}$ containing $M$ subtasks with different task types, a task hypergraph is used to capture the relationship between tasks and the task initiator. The definition of the task hypergraph is as follows.

\begin{definition}[Task Hypergraph]
   \textit{
    	For a task $\bm{B} = \{b_1, \dots, b_M\}$ generated by $a_j$, its task hypergraph is defined as $\mathcal{H}^{\text{ta}} = (\mathcal{V}^{\text{ta}}, \mathcal{E}^{\text{ta}}, \mathcal{A}^{\text{ta}})$, where $\mathcal{V}^{\text{ta}}$ is the set of objects, including all task types in $\bm{B}$ and the task initiator $a_j$, $\mathcal{E}^{\text{ta}}$ represents the set of all task hyperedges, and $\mathcal{A}^{\text{ta}}$ is the attributes of all objects and the weights of all hyperedges. Each task hyperedge $e^{\text{ta}}_{o_s}$ comprises three objects: $a_j$, the task type $o_s$ of $b_m$, and a dummy object $\varnothing$. The data size $d_{b_m}$, processing density $\rho_{b_m}$, and $t^{\text{max}}_{b_m}$ are the attributes of object $o_s$. The weight $w^{\text{ta}}_{o_s}$ of $e^{\text{ta}}_{o_s}$ is the minimum communication rate requirement, which is determined by $\gamma_0$.}
\end{definition}

The generation method of task hypergraph is presented in Algorithm \ref{task-hypergraph}.  For the consistency of notation, all involved objects in $\mathcal{V}^{\text{ta}}$ are reformulated as $\{v_1^{\text{ta}}, \dots, v^{\text{ta}}_{|\mathcal{V}^{\text{ta}}|} \}$, where $|\mathcal{V}^{\text{ta}}|$ is the number of objects. All hyperedges are reformulated as $\{e_1^{\text{ta}}, \dots, e^{\text{ta}}_{|\mathcal{E}^{\text{ta}}|}\}$, where $|\mathcal{E}^{\text{ta}}|$ is the number of hyperedges.

\subsection{Matching Model of Task and Resource Hypergraphs}
\label{matching_task_resource}

According to the collected information from all devices, the collaboration-driven resource hypergraph $\mathcal{H}^{\text{re}}$ is constructed and stored in the BS. Once any task initiator $a_j$ generates a task and send the task-related information to the BS, the BS will execute Algorithm \ref{task-hypergraph} to build the task hypergraph $\mathcal{H}^{\text{ta}}$. Based on the constructed hypergraphs $\mathcal{H}^{\text{re}}$ and $\mathcal{H}^{\text{ta}}$, the problem of resource discovery and allocation in the collaborative computing IoT system is transformed into the problem of matching $\mathcal{H}^{\text{ta}}$ to $\mathcal{H}^{\text{re}}$ to maximize the value of task completion, which actually is the hypergraph matching problem.

\begin{definition}[Hypergraph Matching] 
    \textit{Given two  hypergraphs, the goal of hypergraph matching is to find correspondences between two hypergraphs by considering the affinities of their corresponding vertices and hyperedges such that the matching score is maximized.} 
\end{definition}

 As defined in Subsection \ref{resource hypergraph} and Subsection \ref{task hypergraph}, each resource hyperedge is composed of two devices that can cooperate and a task type, while each task hyperedge includes the task initiator, a subtask, and a dummy node. Therefore, the goal of finding the potential task collaborator for the task initiator is converted into finding a matched resource hyperedge from $\mathcal{H}^{\text{re}}$ for each task hyperedge, as each resource hyperedge indicates the collaborative relationship between two devices. Because both $\mathcal{H}^{\text{re}}$ and $\mathcal{H}^{\text{ta}}$ are constructed as the 3-uniform hypergraph in this research, one of task hyperedges is assumed as $e^{\text{ta}}_{m_1} = (v_{x}^{\text{ta}}, v_{y}^{\text{ta}}, v_{z}^{\text{ta}} ), m_1={1, \dots,|\mathcal{E}^{\text{ta}}|}, x,y,z=1,\dots,|\mathcal{V}^{\text{ta}}|$, and one of resource hyperedges is assumed as $e^{\text{re}}_{m_2} = (v_{x'}^{\text{re}}, v_{y'}^{\text{re}}, v_{z'}^{\text{re}} ), m_2=1,\dots,|\mathcal{E}^{\text{re}}|, x',y',z'=1,\dots,|\mathcal{V}^{\text{re}}|$. A tuple $(f_{xx'},f_{yy'}, f_{zz'})$ is used to represent hyperedge correspondence.  
If $f_{xx'} = 1$, $v^{\text{ta}}_{x}$ matches $v^{\text{re}}_{x'}$, i.e., $v^{\text{ta}}_{x}$ and $v^{\text{re}}_{x'}$ are the same device, e.g. $a_j$, otherwise $f_{xx'} = 0$. Similarly, if $f_{yy'} = 1$, $v_{y}^{\text{ta}}$ matches $v_{y'}^{\text{re}}$, i.e., $v_{y}^{\text{ta}}$ and $v_{y'}^{\text{re}}$ are the same task type, e.g. $o_s$, otherwise $f_{yy'}=0$. If $v^{\text{ta}}_{x}$ and $v^{\text{ta}}_{y}$ match $v^{\text{re}}_{x'}$ and $v^{\text{re}}_{y'}$, respectively, and the weight $w^{\text{re}}_{m_2}$ of $e^{\text{re}}_{m_2}$ is greater than the weight $w^{\text{ta}}_{m_1}$ of $e^{\text{ta}}_{m_1}$, i.e., satisfying the requirement of communication link,  then $v^{\text{re}}_{z'}$ is the potential task collaborator, e.g. $a_i$. Hence, $f_zf_{z'}$ is set to 1. The matching score between $e^{\text{ta}}_{m_1}$ and $e^{\text{re}}_{m_2}$ can be calculated by equation (\ref{value}) using the attributes associated with $e^{\text{ta}}_{m_1}$ and $e^{\text{re}}_{m_2}$. One can observe that the true match between $e^{\text{ta}}_{m_1}$ and $e^{\text{re}}_{m_2}$ is accompanied by a large matching score. 

 \begin{figure}[t!]
	\centering
	\includegraphics[scale=0.76]{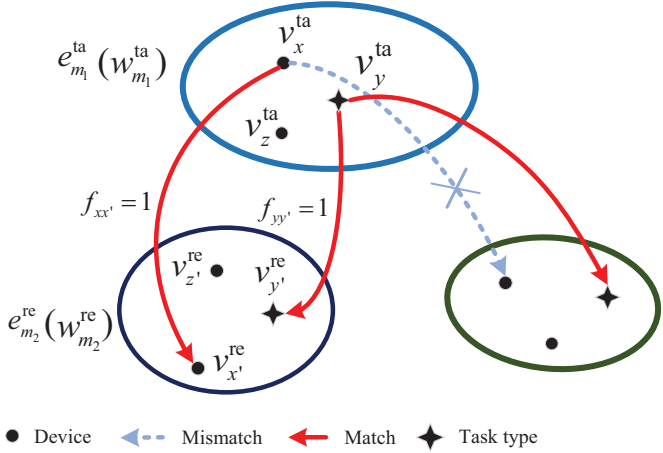}
	\caption{An example of matching between hyperedges.}
	\label{matching}
\end{figure}

Without loss of generality, the number of objects in the resource hypergraph $\mathcal{H}^{\text{re}}$ is much larger than that in the task hypergraph $\mathcal{H}^{\text{ta}}$, i.e., $|\mathcal{V}^{\text{re}}| > |\mathcal{V}^{\text{ta}}|$. To achieve the matching between $\mathcal{H}^{\text{ta}}$ and $\mathcal{H}^{\text{re}}$, the goal is to find a binary assignment matrix $\bm{f} \in \{0, 1\}^{|\mathcal{V}^{\text{ta}}| \times |\mathcal{V}^{\text{re}}|}$. The one-to-one matching constraints are imposed on $\bm{f}$ to ensure that one vertex from $\mathcal{V}^{\text{ta}}$ can match only a single vertex from $\mathcal{V}^{\text{re}}$. 
The set of all potential assignment matrices that map each vertex of $\mathcal{V}^{\text{ta}}$ to each vertex of $\mathcal{V}^{re}$ is written by
\begin{small}
    \begin{align}
    \bm{\mathcal{F}} = \{\bm{f} \in \{0, 1\}^{|\mathcal{V}^{\text{ta}}| \times |\mathcal{V}^{\text{re}}|} | \sum_{x=1}^{|\mathcal{V}^{\text{ta}}|}f_{x,x'} \leq 1, \sum^{|\mathcal{V}^{\text{re}}|}_{x'=1}f_{x,x'} = 1 \}.
\end{align}
\end{small}

For each pair of triples $(v_x^{\text{ta}},v_y^{\text{ta}},v_z^{\text{ta}}) \subset \mathcal{V}^{\text{ta}}$ and $(v^{\text{re}}_{x'}, v^{\text{re}}_{y'}, v^{\text{re}}_{z'}) \subset \mathcal{V}^{\text{re}}$, the matching score is calculated by the matching function $F_{xx',yy',zz'}$. The similarity of two hypergraphs is typically used as the matching score in the classical hypergraph matching problem. However, in this research, the value of task completion is utilized as the matching score to evaluate the degree of matching between two hypergraphs. Therefore, the overall matching score $S^{\text{score}}(\bm{f}), \bm{f} \in \bm{\mathcal{F}}$ is defined as follows
\begin{align}
   \label{score}
    S^{\text{score}} (\bm{f}) = \sum_{x,y,z=1}^{|\mathcal{V}^{\text{ta}}|} \sum_{x',y',z'=1}^{|\mathcal{V}^{\text{re}}|} F_{xx',yy',zz'}f_{x,x'}f_{y,y'}f_{zz'}.
\end{align}
The product $f_{x,x'}f_{y,y'}f_{zz'}$ is equal to 1 if and only if the triples $(v_x^{\text{ta}},v_y^{\text{ta}},v_z^{\text{ta}})$ are matched to $(v^{\text{re}}_{x'}, v^{\text{re}}_{y'}, v^{\text{re}}_{z'})$, respectively. Then, $ F_{xx',yy',zz'}$ is added to the total matching score.  

To find the optimal matching that maximizes the overall matching score, the hypergraph matching problem between $\mathcal{H}^{\text{ta}}$ and $\mathcal{H}^{\text{re}}$ is formulated as
\begin{alignat}{1}
    P2: \bm{f}^{*} = \argmax\limits_{\bm{f} \in \bm{\mathcal{F}}} \quad S^{\text{score}}(\bm{f}).
\end{alignat}
One can see the optimization problem $P1$ that jointly optimizes resource selection and allocation is successfully converted into the hypergraph matching problem $P2$, which constitutes a matching problem encompassing higher-order collaborative relationships among computing, communication, and tasks, rather than a simple point-to-point matching. As a result, conventional graph matching methods are unsuitable for addressing the problem in this study. Therefore, a game-theoretic hypergraph matching method is proposed to achieve this matching.

\section{Game-Theoretic Hypergraph Matching}
\label{game-theoretic matching}

\begin{figure*}[t!]
	\centering
	\includegraphics[scale=0.64]{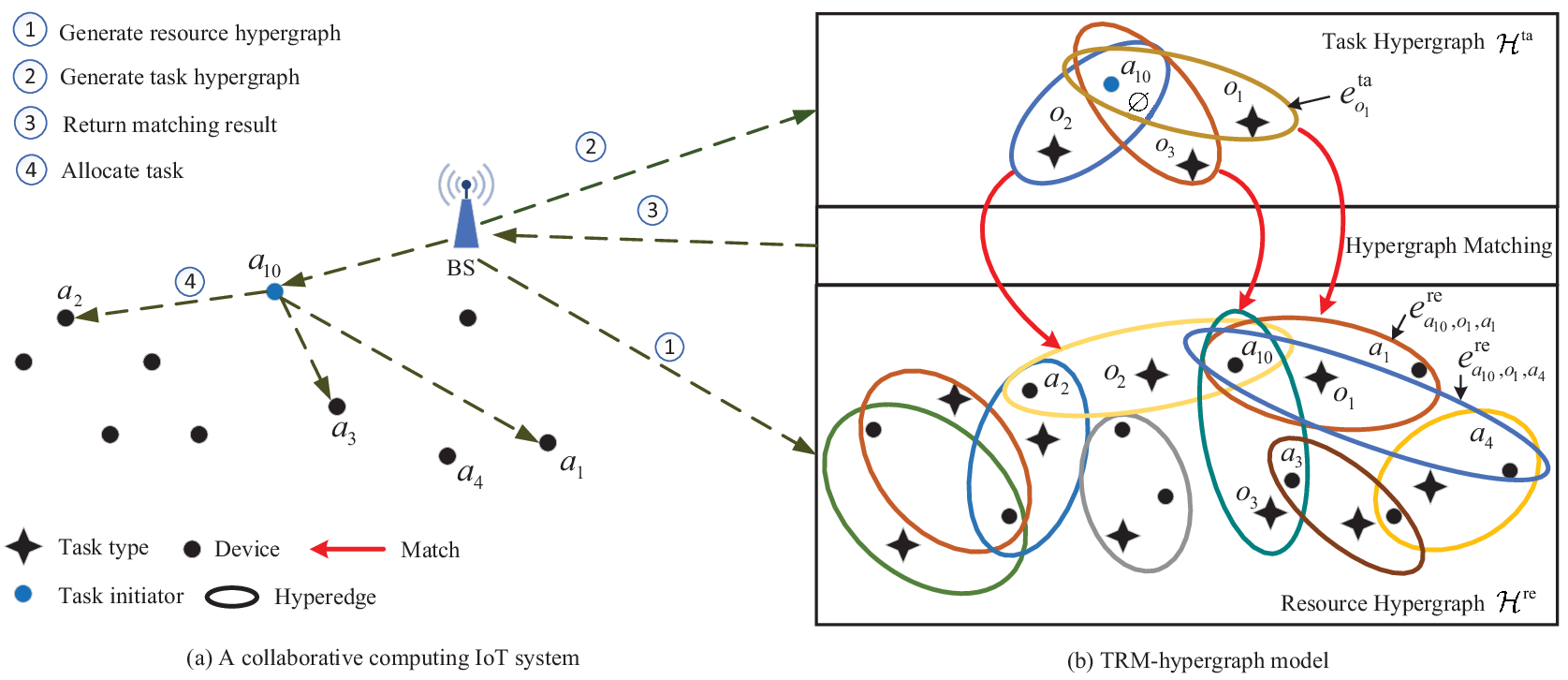}
	\caption{TRM-hypergraph model-driven tasks and resources matching in the collaborative computing IoT system. }
	\label{hypercyber}
\end{figure*}

Existing hypergraph matching algorithms, e.g., \cite{4587500}, \cite{5871640}, \cite{5432196}, use the tensor-based algorithms to solve the matching problem, leading to high computational loads. In this research, we find that if a task hyperedge $e^{\text{ta}}_{m_1} \in \mathcal{H}^{\text{ta}}$ is truly matched with a resource hyperedge $e^{\text{re}}_{m_2} \in \mathcal{H}^{\text{re}}$, this matching will be associated with a high matching score, i.e., value of task completion, whereas if they are falsely matched, this matching will be associated with a low matching score. This observation suggests that a set of correct matches forms a coherent group, supporting each other to maximize the overall matching score. In contrast, a set of false matches is unlikely to be coherent with each other due to the small matching scores. Likewise, the mixture of both correct and false matches cannot form a coherent group. Thus, we can regard a set of correct matches as a cluster when utilizing all matches as data for clustering. This insight leads us to transform the hypergraph matching problem into a non-cooperative multiplayer clustering game. The detailed discussion of addressing the hypergraph matching problem via the clustering game is presented below.

\subsection{Hypergraph Matching as a Clustering Game}

In terms of game theory \cite{8545827}, a game of strategy is formalized as $\Gamma = (\bm{P}, \bm{N}, \pi)$, where $\bm{P}=\{1,\dots, P\}$ is the set of players, $\bm{N}$ is the set of pure strategies that are available to players. $\pi: \bm{N}^{P} \rightarrow \mathbb{R}$ is the payoff function that assigns a payoff to each strategy profile defined as an ordered group of strategies played by the different players. All candidate matches between $\mathcal{H}^{\text{ta}}$ and $\mathcal{H}^{\text{re}}$ are treated as the pure strategies, and matching score values are considered as the payoffs of players. 
 
The clustering game is played in an evolutionary scenario in which players are drawn from a large population to repeatedly play the game. All players are not supported to behave rationally or have a complete knowledge of the game details. In the game, each player is assumed to play a predefined strategy and receives an associated payoff. According to the Darwinian Natural Selection theory, players who achieve significant payoffs thrive over the course of evolution, whereas those with lower payoffs progressively diminish. As payoffs are evaluated by the matching score values, it becomes evident that the survived strategies (matches) played by the survived players receive high matching scores, whereas the extinct ones are linked with minor matching scores. Hence, the survived matches can be seen as a cluster. This cluster can be further found by obtaining an evolutionary stably strategy (ESS) cluster of the clustering game.

\subsection{Game-theoretic Hypergraph Matching Algorithm}

To get the final cluster including correct matches, the proposed game-theoretic hypergraph matching algorithm is implemented in two steps: finding candidate matches and selecting the ESS cluster.

\subsubsection{Finding candidate matches}
 
For each task hyperedge $e^{\text{ta}}_{m_1}$ in the task hypergraph $\mathcal{H}^{\text{ta}}$, multiple coarse correspondences between $e^{\text{ta}}_{m_1}$ and resource hyperedges in the resource hypergraph $\mathcal{H}^{\text{re}}$ are established. According to the coarse correspondences of task hyperedges and resource hyperedges, the set of candidate matches between $\mathcal{H}^{\text{ta}}$ and $\mathcal{H}^{\text{re}}$ is obtained, denoted as $\bm{N}=\{1, \dots, N\}$, which includes all strategies.    

\subsubsection{Selecting the ESS-cluster}
According to \cite{6330964}, the ESS-cluster of a hypergraph clustering problem includes two basic properties: internal coherency and external incoherency, which reveals the fact that elements in the ESS cluster should have high mutual similarities. The ESS cluster corresponds to the equilibria of the clustering game, which can be iteratively derived by optimizing a polynomial function over the standard simplex.

Given the set of strategies $\bm{N}$ and three players, $P=3$, the set of all states of the population is represented by the standard simplex $\Delta$ of $N$-dimensional Euclidean space, which is denoted as
\begin{align}
   \hspace{-2.5mm} \Delta = \{ \bm{q} \in \mathbb{R}^{N}:\sum_{n = 1}^{N} q_n = 1 \  \text{and} \  q_n \geq 0 \ \text{for all} \ n \in \bm{N} \}, 
\end{align}
where $\bm{q}$ is the state of the population at a given time $t$, and $q_n$ represents the fraction of players selecting the $n$-th strategy at time $t$. As time passes, the state of the population evolves through the impact of a selection process. Eventually, this state reaches an equilibrium described by a Nash equilibrium, i.e., $\bm{q} \in \Delta$ is a Nash equilibrium if 
\begin{align}
   \label{nash}
    u \left( \bm{\nu}^{n}, \bm{q}^{[2]} \right) \leq u \left(\bm{q}^{[3]} \right), \ \text{for all} \ n \in \bm{N}, 
\end{align}
where the left side stands for the expected payoff earned by players selecting the $n$-th strategy, and the expected payoff over the entire population is given by the right side. $\bm{\nu}^{n}$ defines a $N$-dimensional state vector with $q_n = 1$ and zero elsewhere. $\bm{q}^{[3]}$ stands for 3 identical state $\bm{q}$. 
One can see that each player acts at most as well as the entire population expected payoff.
Since the survived matches are regarded as a cluster, this cluster can be obtained by identifying a Nash equilibrium. However, it is found that the Nash equilibrium lacks stability concerning small perturbations in a population-based setting. The authors in \cite{Smith_1982} introduced ESS as the refinement of the Nash equilibrium, and the cluster of survived matches can be obtained by finding out the ESS from the clustering game. Hence, by using the same idea with \cite{Smith_1982}, the Baum-Eagon inequality is utilized to offer a practical iterative approach to extract the ESS-cluster \cite{6330964}
\begin{align}
    \label{ess}
    q_n(t+1) = q_n(t) \frac{ u \left( \bm{\nu}^{n}, \bm{q}(t)^{[2]} \right)}{u\left(\bm{q}(t)^{[3]} \right)}, n=1,\dots,N,
\end{align}
and the function $u()$ is given by
\begin{align}
    u\left(\bm{h}^{(1)},\bm{h}^{(2)}, \bm{h}^{(3)}\right) = \sum_{\substack{ (n_1,n_2,n_3) \in \bm{N}^{3}}} \pi \left(n_1,n_2,n_3 \right) \prod_{p=1}^{3} h^{(p)}_{n_p},
\end{align}
where $\pi (n_1,n_2,n_3)$ is the assigned payoff, i.e., value of task completion, to the strategy profile $(n_1,n_2,n_3) \in \bm{N}^3$ and $n_p$, $p=1,2,3$, is the strategy selected by the $p$-th player. According to the analysis in Subsection \ref{matching_task_resource}, if all three strategies in $(n_1,n_2,n_3)$ are true matches, $\pi(n_1,n_2,n_3)$ will receive a high payoff, indicating a high matching score or a high value of task completion. At time 0, the dynamic starts from an initial state $\bm{q}(0)$ and iteratively updates the state until the final state $\bm{q}*$ is obtained at convergence. Since the entries of $\bm{q}$ reflect the proportions of strategies in the population state \cite{8545827}, we refer to entries as the weights of the corresponding strategies. The survived strategies with weights above a threshold in $\bm{q}*$ constitute a cluster $\bm{C}$ in which the matches between $\mathcal{H}^{\text{ta}}$ and $\mathcal{H}^{\text{re}}$ will be regarded as the correct matches. As each match within the ESS cluster is assigned a weight, denoting its degree of alignment with other matches, a higher weight indicates a greater likelihood of the match being accurate. Once one-to-many matches appear in the cluster, the match with the largest weight is preserved for following the one-to-one constraint. Finally, the ESS cluster $\bm{C}^*$ with the correct matches that can ensure the maximal value of task completion is obtained. The detailed implementation of the game-theoretic hypergraph matching algorithm is presented in Algorithm \ref{game matching}. The proposed TRM-hypergraph model that transforms the problem of discovering and allocating computing and communication resources into the matching between the task hypergraph $\mathcal{H}^{\text{ta}}$ and the resource hypergraph $\mathcal{H}^{\text{re}}$ is illustrated in Fig.~\ref{hypercyber}. The operational workflow of the entire model is summarized as follows. 
After receiving information from all devices, the BS creates a resource hypergraph. 
When device $a_{10}$ as the task initiator generates a new task consisting of subtasks of three different types ($o_1$, $o_2$, and $o_3$) and requests resources from the BS, the BS will generate a task hypergraph. Then, the matching between the task hypergraph and the resource hypergraph is performed using the game-theoretic hypergraph matching method. As shown in Fig.~\ref{hypercyber}, although both resource hyperedges $e^{\text{re}}_{a_{10},o_1,a_1}$ and $e^{\text{re}}_{a_{10},o_1,a_4}$ are matched with the task hyperedge $e^{\text{ta}}_{o_1}$, $e^{\text{re}}_{a_{10},o_1,a_1}$ and $e^{\text{ta}}_{o_1}$ are ultimately the most matched through the iterative clustering game. Hence, device $a_1$ is selected to execute the subtask of task type $o_1$. Finally, $a_1$, $a_2$, and $a_3$ are the most suitable devices to execute the subtasks from $a_{10}$.

\begin{algorithm}[t!]
	\caption{{Game-theoretic hypergraph matching algorithm}}
	\label{game matching}
	\begin{algorithmic}[1]
		\renewcommand{\algorithmicrequire}{\textbf{Input:}}
		\REQUIRE $\mathcal{H}^{\text{ta}}, \mathcal{H}^{\text{re}}$,  maximum number of iterations $T$
		\renewcommand{\algorithmicrequire}{\textbf{Output:}}
		\REQUIRE $\bm{C}^{*}$ 
		\FOR{each $e^{\text{ta}}_{m_1}$ in $\mathcal{H}^{\text{ta}}$}
		\STATE coarse matches between $e^{\text{ta}}_{m_1}$ and $ \mathcal{H^{\text{re}}}$
		\ENDFOR
		\STATE $\bm{N}=\{1,\dots,N\}$
		\STATE $t \leftarrow 0$ \\
		\STATE $\bm{q}^{*} \leftarrow \bm{q}(0)$\\
		\WHILE{$t < T$}
		\STATE $t \leftarrow t + 1$ \\
		\FOR{each $n$ in $\bm{N}$}
		\STATE update $q_n(t)$ by equation (\ref{ess})\\
		\ENDFOR
		\STATE $\bm{q}^{*} \leftarrow \bm{q}(t)$\\
		\ENDWHILE
		\STATE $\bm{C} \xleftarrow[]{}{\bm{q}^{*}}$ for weights above threshold\\
		\STATE $\bm{C}^{*} \xleftarrow[]{one-to-one} {\bm{C}}$
		\RETURN  $\bm{C}^{*}$\\
	\end{algorithmic}
\end{algorithm}

\subsection{Computational Complexity Analysis}
The computational complexity of generating $\mathcal{H}^{\text{re}}$ is $O(J^2 S)$, where $J$ is the number of all devices and $S$ is the number of task types. In addition, the computational complexity in Algorithm \ref{task-hypergraph} for generating $\mathcal{H}^{\text{ta}}$ is $O(M)$, where $M$ is the number of subtasks. Furthermore, the computational complexity of game-theoretic hypergraph matching is at most $O(NT + N^2)$~\cite{HOU2023109035}, where $N$ is the number of all strategies, $T$ denotes the number of iterations, and $N^2$ is performing the one-to-one constraint. Therefore, the computational complexity of the whole algorithm is $O(J^2 S + M + NT + N^2)$.

\section{Simulation Results}
\label{section simulation}

 \begin{figure}[!t]
      \centering
      \subfigure[$\xi_1 = 0$, $\xi_2 = 1$.]{\includegraphics[scale=0.66]{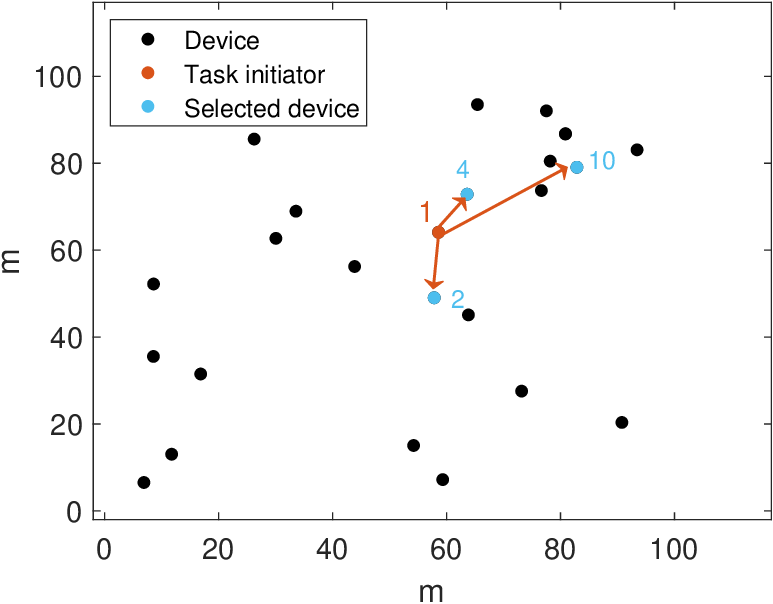}}
	  \subfigure[$\xi_1 = 1$, $\xi_2 = 0$.]{\includegraphics[scale=0.66]{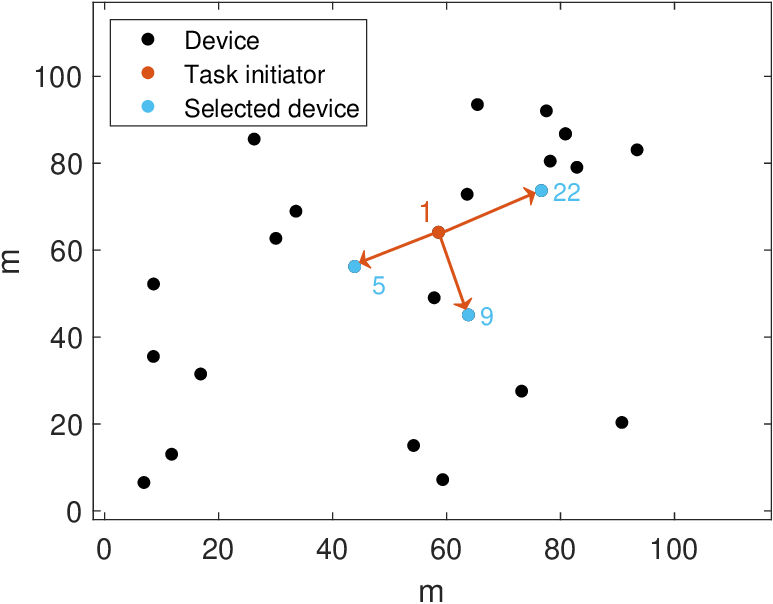}}
     \caption{Selected collaborative computing devices.}
     \label{selected}
 \end{figure}
   
In this section, the proposed TRM-hypergraph method is evaluated via numerical and simulation results using Matlab. The simulation parameters are set as follows. The considered scenario is in $100$ m $\times$ $100$ m area where $J=25$ devices are randomly scattered and the BS is located in the center of the region. For the wireless access, the channel bandwidth $W$ is set to 5 MHz, and the noise $N_0 = -100$ dBm. The transmission power of each device is randomly assigned  from the set $\{100, 150,200\}$ mW. For the computational task, the data size of each subtask is randomly generated from the interval $[200, 1000]$ KB. The total number of task types is assumed as $S=5$, hence, $\bm{O} = \{o_1, o_2, o_3, o_4, o_5\}$. Each subtask corresponds to a different task type whose processing density is randomly assigned from the set $\{500, 600, 700, 800\}$ cycles/bit. For the device resource, each device's ID is assigned a non-repeating number from $\{1,2,\dots, 25\}$ during initialization, and the states of all devices are set to idle by default. For each device, its CPU clock frequency is randomly selected from the set $\{0.5, 0.8, 1 \}$ GHz, and it is also assumed to support three different task types selected from $\bm{O}$. It is worth pointing out that the results in the following subsections are the average results of 100 times.

\begin{table}[t!]
	\centering
        \renewcommand{\arraystretch}{1.3}
	\caption{Task features generated by the task initiator}
	\label{table1}
	\footnotesize
	  \begin{tabular}{p{1.3cm}<{\centering}|p{1.3cm}<{\centering}|p{1.3cm}<{\centering}|p{3cm}<{\centering}}
		\hline \textbf{Subtask} & \textbf{Task type} & \textbf{Size (KB)} & \textbf{Processing density (cycles/bit)} \\
		\hline\hline
		1 & $o_1$ & 722 & 600\\
		\hline
		2 & $o_4$ & 272 & 500\\
		\hline
		3 & $o_5$ & 861 & 400\\
		\hline
   \end{tabular}
\end{table}

\begin{table}[t]
	\centering
        \renewcommand{\arraystretch}{1.3}
	\caption{Selected devices in different values of $\xi_1$ and $\xi_2$}
	\label{table2}
	  \begin{tabular}{p{1.9cm}<{\centering}|p{1.6cm}<{\centering}|p{1.6cm}<{\centering}|p{1.6cm}<{\centering}}
            \hline \multirow{2}*{} & \multicolumn{3}{c}{\textbf{Selected devices}} \\
            \cline{2-4} & $o_1$ & $o_4$ & $o_5$\\
            \hline\hline
		$\xi_1 = 0, \xi_2 = 1$  & 4 (0.5 GHz) & 10 (0.5 GHz) & 2 (0.5 GHz)\\
		\hline
		$\xi_1 = 1, \xi_2 = 0$  & 22 (0.8 GHz) & 9 (1 GHz) & 5 (1 GHz)\\
		\hline
   \end{tabular}
\end{table}

 \begin{figure*}[!t]
      \centering
      \subfigure[$\xi_1 = 0$, $ \xi_2 = 1$.]{\includegraphics[scale=0.449]{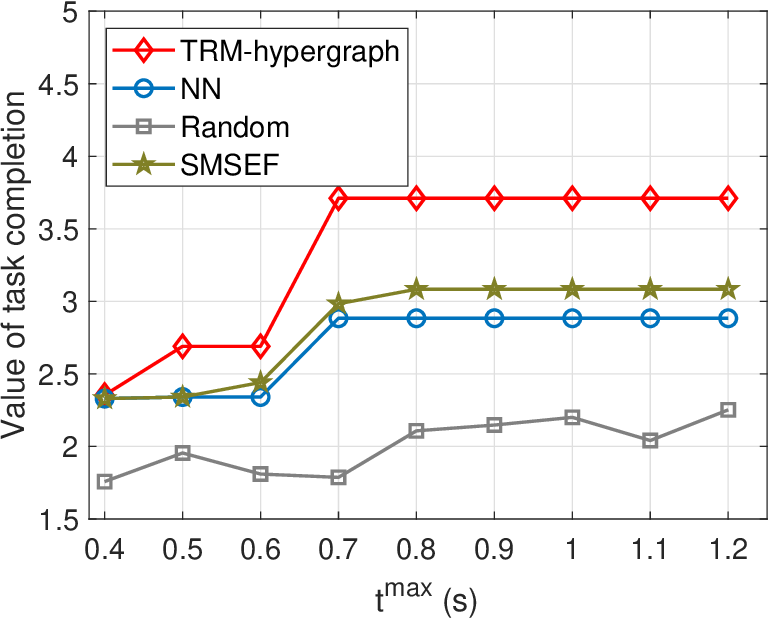}}\hspace{0.1in}%
   \subfigure[$\xi_1 = 1$, $\xi_2 = 0$.]{\includegraphics[scale=0.449]{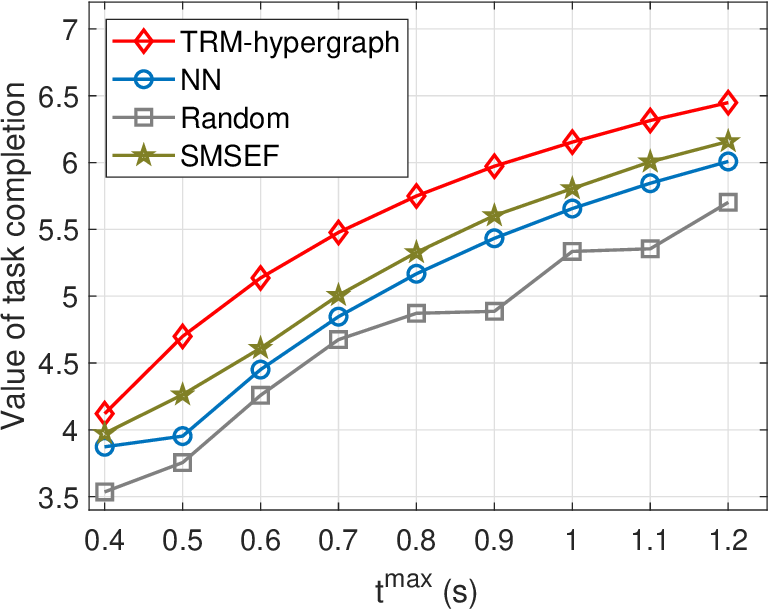}}\hspace{0.1in}%
    \subfigure[$\xi_1 = 0.5$, $\xi_2 = 0.5$.]{\includegraphics[scale=0.449]{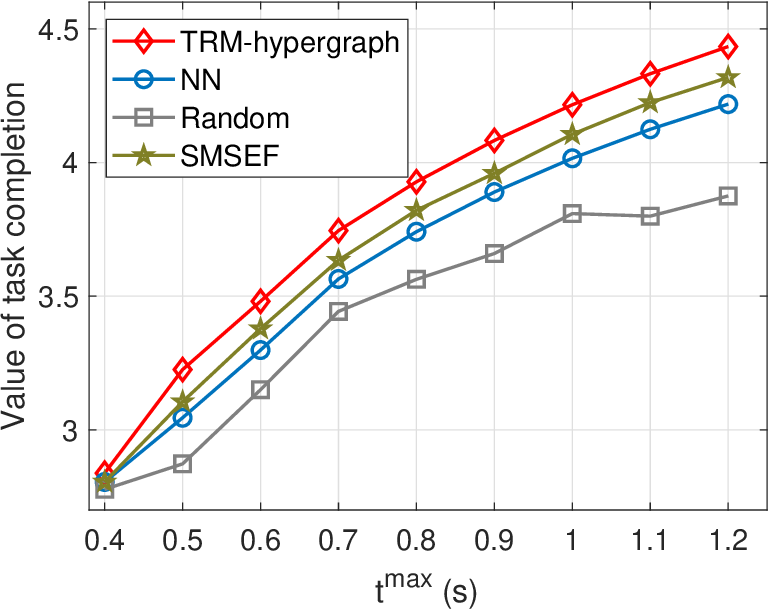}}
     \caption{Comparisons of value of task completion.}
     \label{deadline}
\end{figure*}

\subsection{Effect of $\xi_1$ and $\xi_2$ on Selecting Devices}

 \begin{figure*}[!t]
      \centering
      \subfigure[$t^{\text{max}} = 0.4 $ s.]{\includegraphics[scale=0.455]{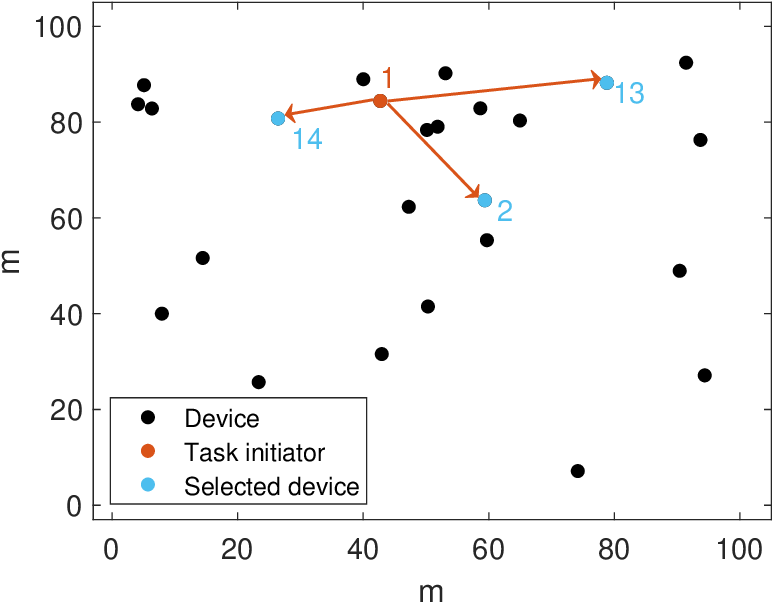}}\hspace{0.05in}%
   \subfigure[$t^{\text{max}} = 0.5 $ s.]{\includegraphics[scale=0.455]{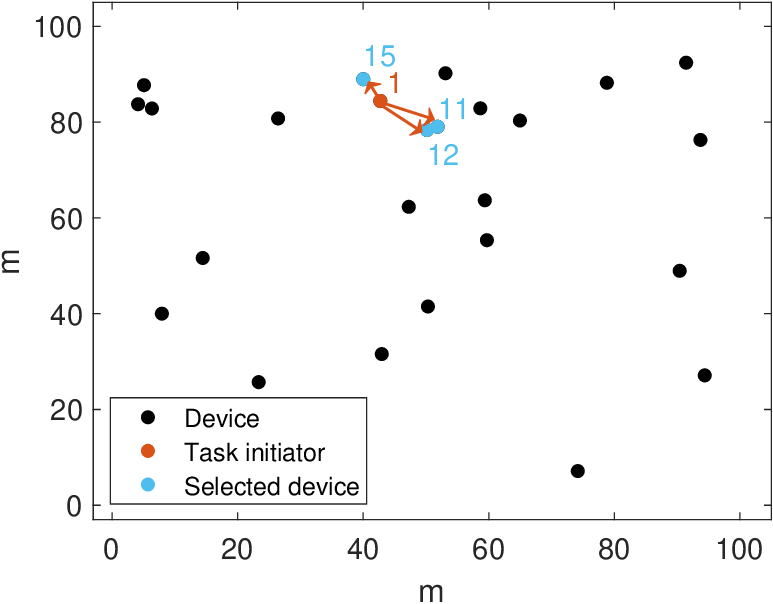}}\hspace{0.05in}%
    \subfigure[$t^{\text{max}} = 0.7 $ s.]{\includegraphics[scale=0.455]{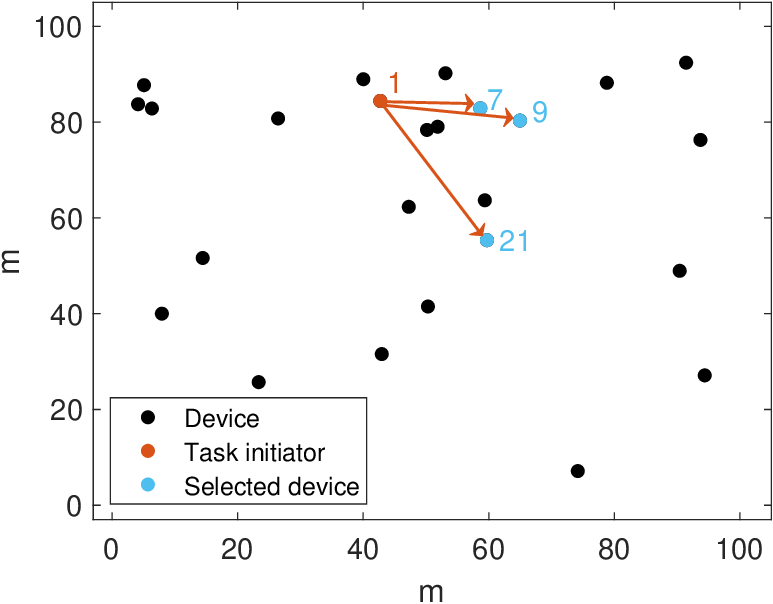}}
     \caption{Selected collaborative computing devices by TRM-hypergraph when $\xi_1 = 0$ and $\xi_2 = 1$.}
     \label{selected deadline}
\end{figure*}

In this subsection, the effect of the choices of $\xi_1, \xi_2$ on the selection of collaborative computing devices is assessed. According to (\ref{value}), the value of task completion includes only the value of energy when $\xi_1 = 0$ and $\xi_2 = 1$. On the contrary, the value of task completion consists solely of the value of energy if $\xi_1 = 1$ and $\xi_2 = 0$. Device 1 is selected as the task initiator which includes three different subtasks with types $o_1$, $o_4$, and $o_5$. The details of subtasks are listed in Table \ref{table1}, and the deadline for all subtasks is set to 0.9~s. Take the subtask 3 as an example. One can see that device 2 is selected as the collaborative computing device to execute the subtask 3 when $\xi_1 = 0$ and $\xi_2 = 1$, as shown in Fig.~\ref{selected} (a) and Table~\ref{table2}. This is not only because device 2 supports the task type $o_5$, but also because device 2 is closer to device 1 and has a lower  CPU clock frequency (0.5 GHz), which can significantly reduce the energy consumption in transmission and computation. This is because device 2 not only supports the task type $o_5$, but also it has a short distance to device 1 and a lower CPU clock frequency (0.5 GHz). Therefore, device 1 can receive the high value of task completion from completing the subtask 3.

When $\xi_1 = 1, \xi_2 = 0$, device 5 is selected to execute the subtask 3, as shown in Fig.~\ref{selected}~(b) and Table~\ref{table2}. In this case, device 5 can quickly complete the subtask 3 because it has a high CPU clock frequency (1 GHz). Hence, device 1 spends very little waiting time to receive the calculation result from device 5, which leads to the high value of task completion. From the aforementioned analysis, it becomes evident that the selections of $\xi_1$ and $\xi_2$ exert a significant influence on the outcomes of collaborative computing device selection. But no matter in which case, the appropriate devices are always selected to perform the corresponding subtasks by our proposed TRM-hypergraph algorithm.

\subsection{Impact of $t^{\text{max}}$ on Value of Task Completion}

\begin{table}[t]
	\centering
        \renewcommand{\arraystretch}{1.3}
	\caption{Frequencies of selected devices in different $t^{\text{max}}$}
	\label{}
	  \begin{tabular}{p{1.9cm}<{\centering}|p{1.6cm}<{\centering}|p{1.6cm}<{\centering}|p{1.6cm}<{\centering}}
            \hline \multirow{2}*{$t^{\text{max}}$ (second)} & \multicolumn{3}{c}{\textbf{Selected devices}} \\
            \cline{2-4} & $o_2$ & $o_3$ & $o_4$\\
            \hline\hline
		0.4  & 14 (1 GHz) & 13 (1 GHz) & 2 (1 GHz) \\
		\hline
	    0.5  & 12 (0.8 GHz)  & 15 (0.8 GHz)  & 11 (0.8 GHz) \\
            \hline
            0.7  &  9 (0.5 GHz) & 7 (0.5 GHz)  & 21 (0.5 GHz)  \\
		\hline
   \end{tabular}
\end{table}

 \begin{figure}[!t]
      \centering
      \subfigure[Comparisons of total energy consumption.]{\includegraphics[scale=0.6]{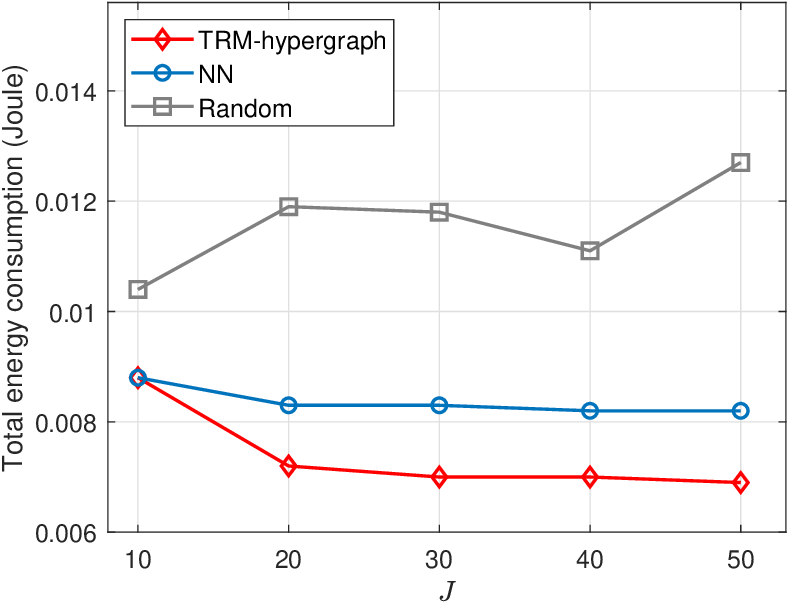}}
    \subfigure[ Comparisons of value of task completion.]{\includegraphics[scale=0.61]{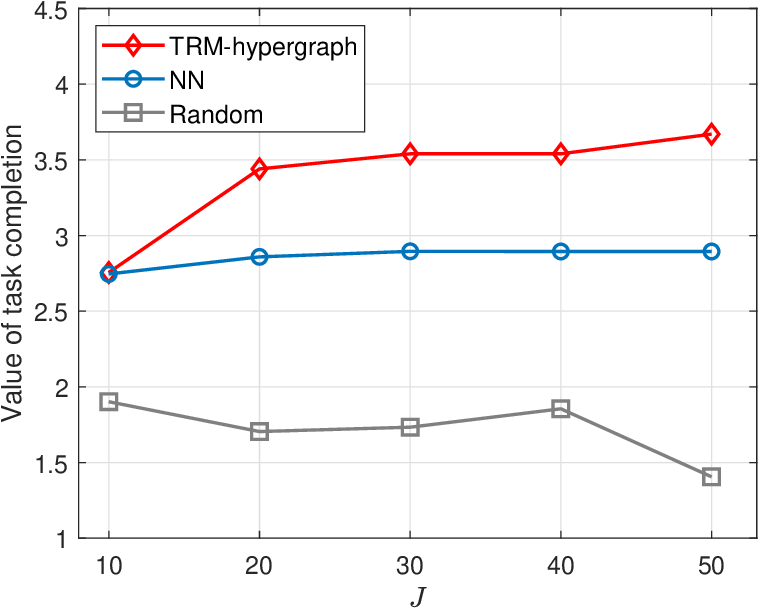}}
     \caption{Comparisons of value of task completion and energy consumption when varying the number of devices in $\xi_1 = 0$ and $\xi_2 = 1$.}
     \label{comparison energy}
\end{figure}

In this subsection, the influence of the maximal task completion time on the value of task completion is assessed by conducting a comparative analysis between our proposed TRM-hypergraph method and three other algorithms: nearest neighbor (NN) search, random search, and SMSEF \cite{8543056}. The task generated by device 1 includes three subtasks with types $o_2, o_3, o_4$. Fig.~\ref{deadline} illustrates the value of task completion comparisons when varying task completion times. One can observe that the value of task completion almost always increases with the increase of $t^{\text{max}}$, no matter which weight parameters are used. When $t^{\text{max}}$ is relatively small, the number of selectable devices is limited, which can result in high task completion time or increased energy consumption, ultimately leading to a lower value of task completion. As $t^{\text{max}}$ increases, the number of devices that can be selected is increased, and more suitable devices are chosen to perform the subtasks to obtain the high value of task completion. The proposed TRM-hypergraph algorithm consistently outperforms three comparison algorithms in achieving the high value of task completion.

As shown in Fig.~\ref{selected deadline}~(a), when $t^{\text{max}} = 0.4$ second, devices 14, 13, and 2 are selected to perform subtasks $o_2$, $o_3$, and $o_4$, respectively. This is because these three devices have the highest CPU clock frequency (1 GHz), which can ensure all subtasks are completed within 0.4 second. However, these three devices consume more energy to process the subtasks, which results in a low value of task completion, i.e., 2.3543. As shown in Fig. \ref{selected deadline}~(b), devices 11, 12, and 15 with medium CPU clock frequency are selected to execute the subtasks when $t^{\text{max}} = 0.5$ second, and the value of task completion increases to 2.6902. When $t^{\text{max}} = 0.7$ second, devices 7, 9, and 21 are selected to process the subtasks. The obtained value of task completion is increased to 3.7107 because of less computational energy consumption.

\begin{table}[t]
	\centering
        \renewcommand{\arraystretch}{1.3}
	\caption{Comparison of running time}
	\label{running_time}
	  \begin{tabular}{p{1cm}<{\centering}|p{1.2cm}<{\centering}|p{1.2cm}<{\centering}|p{1.2cm}<{\centering}|p{2.1cm}<{\centering}}
            \hline
		 & {NN} & {Random} & {SMSEF} &  {TRM-hypergraph} \\
		\hline \hline
		{Time (s)} & {0.8834} & {0.1410} & {1.163} & {0.6920} \\ 
		\hline
   \end{tabular}
\end{table}

The average running time of the algorithms is further compared by executing 500 tasks. As shown in Table \ref{running_time}, the proposed TRM-hypergraph algorithm exhibits lower running time compared to NN and SMSEF. This is because when the number of devices in the system remains unchanged, the TRM-hypergraph algorithm only spends time creating a resource hypergraph in the first task. In each subsequent task, its running time mainly includes the execution of the matching algorithm. However, NN and SMSEF need to execute all algorithm-related procedures for each task, which may result in relatively high running time.

\subsection{Impact of Varying the Number of Devices}

 \begin{figure}[!t]
      \centering
      \subfigure[Comparisons of total task completion time.]{\includegraphics[scale=0.61]{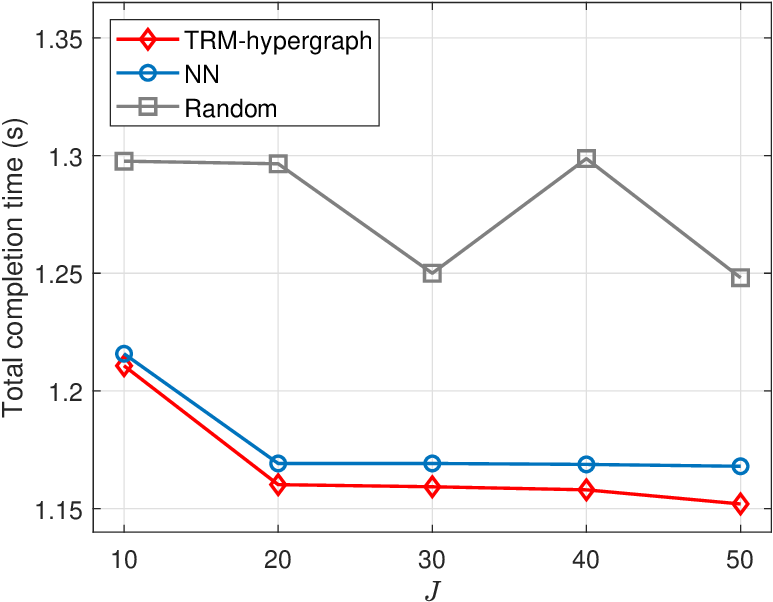}}
    \subfigure[ Comparisons of value of task completion.]{\includegraphics[scale=0.61]{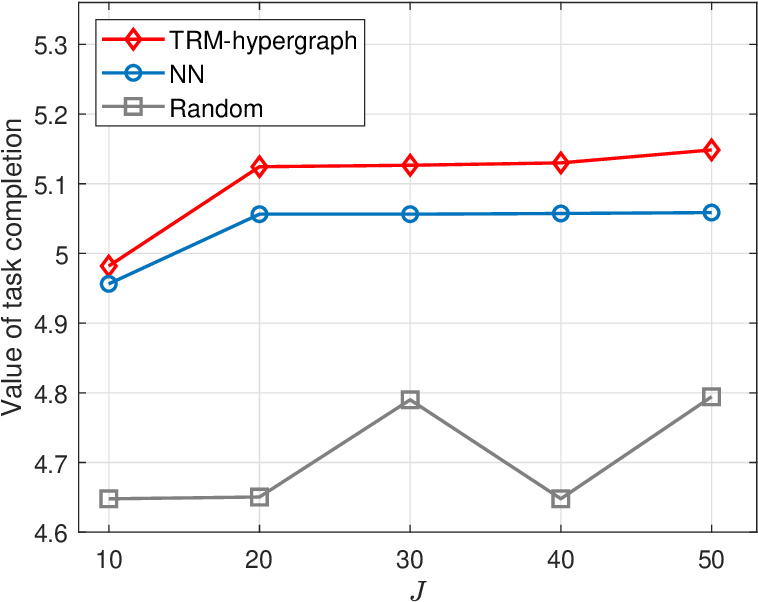}}
     \caption{Comparisons of value of task completion and time when varying the number of devices in $\xi_1 = 1$ and $\xi_2 = 0$.}
     \label{comparison time}
\end{figure}

In this subsection, the impact of varying the number of devices on task is investigated when $t^{\text{max} }= 0.8$ second. As illustrated in Fig.~\ref{comparison energy} (a), the total energy consumption of our proposed TRM-hypergraph method decreases as the number of devices ($J$) increases when $\xi_1 = 0$ and $\xi_2 = 1$. The increase in available devices results in the task initiator being able to select the optimal matching devices spending the least amount of energy consumption in computation and transmission. Hence, these selected devices achieve the optimal value of task completion in comparison to the NN and random algorithms, as shown in Fig.~\ref{comparison energy} (b).

In Fig.~\ref{comparison time} (a), as the number of devices increases, the total task completion time obtained by our proposed TRM-hypergraph is decreased as compared to the NN and random search algorithms. Thus, the proposed TRM-hypergraph outperforms comparison algorithms in achieving the maximal value of task completion, as shown in Fig.~\ref{comparison time} (b).

\subsection{Assessing the Performance of TRM-hypergraph in the Real World}

 \begin{figure}[!t]
      \centering
      \subfigure[A swarm robot system.]{\includegraphics[scale=0.51]{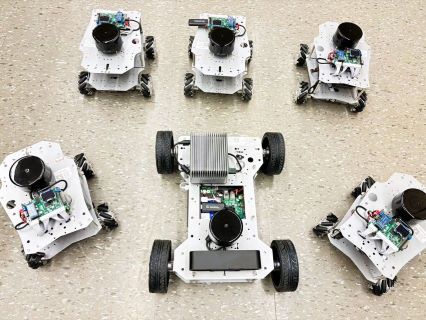}}
    \subfigure[Comparisons of value of task completion.]{\includegraphics[scale=0.61]{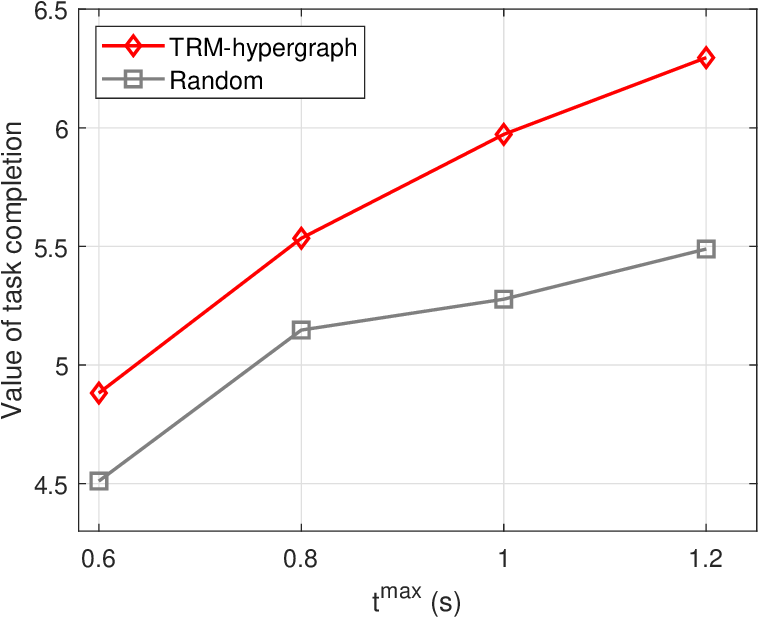}}
     \caption{Comparisons of value of task completion in a swarm robot system.}
     \label{robot}
\end{figure}

In this subsection, a Rosbot Plus, acting as the task initiator, and 5 Rosbots (CPU: ARM M4, 168 MHz), serving as the collaborative devices, are used to model a collaborative computing system to evaluate the performance of the proposed TRM-hypergraph model. The task initiator generates three different tasks, each of which is a code program with a certain number of loops.
By setting $\xi_1 = 1$ and $\xi_2 = 0$, the change in value of task completion is observed as $t^{\text{max}}$ varies. As shown in Fig.~\ref{robot}~(b), the TRM-hypergraph consistently achieves the highest value of task completion compared to the baseline algorithm in a collaborative computing system simulated by a swarm robot system.

\section{Conclusion}
\label{conclusion}
In this paper, the state-of-the-art TRM-hypergraph model is proposed to address the problem of resource selection and allocation in the collaborative computing IoT system. Initially, the resource hypergraph and the task hypergraph are presented to accurately build the relationships among complex tasks, intricate computing resources, and communication resources. Then, the challenge of resource selection and allocation is converted into the hypergraph matching problem between the task hypergraph and the resource hypergraph. Furthermore, the game-theoretic approach is designed to solve the hypergraph matching problem by viewing this problem as the non-cooperative multiplayer clustering game. Numerical results demonstrate that the proposed TRM-hypergraph model achieves the maximal value of task completion under different weight parameters when compared to other benchmark methods. 

\footnotesize
\bibliographystyle{IEEEtranN}

\bibliography{IEEEabrv, bibliography_file.bib} 


\end{document}